\documentclass[12pt,a4paper,twoside]{article}
\usepackage[utf8]{inputenc}
\usepackage[T1]{fontenc}
\usepackage{hyphsubst}
\usepackage[margin=2.5cm]{geometry}
\usepackage{lmodern}
\usepackage[square, comma, numbers, sort&compress, round]{natbib}
\usepackage[english]{babel}
\usepackage{mathtools, nccmath}
\usepackage{bef_aras_nothm,todonotes}
\usepackage{amsmath,amsthm,amssymb}
\usepackage{fixmath}
\usepackage{upgreek}
\usepackage{amsfonts}
\usepackage{microtype}
\usepackage{float}
\usepackage{mathrsfs}
\usepackage{hyperref}
\usepackage{paralist}
\hypersetup{
     colorlinks   = true,
     citecolor    = blue
}
\usepackage[nameinlink]{cleveref} 
\crefname{figure}{\figurename}{\figurename}  
\usepackage{tikz}
\usepackage{verbatim}
\usepackage{afterpage}
\usepackage{glossaries}

\usepackage[headsepline]{scrlayer-scrpage}
\clearpairofpagestyles
\automark[subsection]{section}

\usetikzlibrary{arrows,calc}

\tikzset{
>=stealth',
help lines/.style={dashed, thick},
axis/.style={<->},
important line/.style={thick},
connection/.style={thick, dotted},
}

\theoremstyle{plain}
\newtheorem{theorem}{Theorem}[section]

\newtheorem{pro}[theorem]{Proposition}
\newtheorem{defi}[theorem]{Definition}
\newtheorem{cor}[theorem]{Corollary}
\theoremstyle{definition}
\newtheorem{con}[theorem]{Conjecture}

\newtheorem{rem}[theorem]{Remark}

\newcounter{ArasCounter}

\newcounter{GittaCounter}

\newcounter{HolgerCounter}

\makeatletter
\g@addto@macro{\thm@space@setup}{\thm@headpunct{}}
\makeatother

\numberwithin{equation}{section}

\allowdisplaybreaks

\begin{document}

\newpage
\setcounter{page}{1}
\clearpairofpagestyles
\ohead{\pagemark}
\ihead{\headmark}

\title{Complexity Blowup for Solutions of the Laplace and the Diffusion Equation}

\author{Aras Bacho\footnotemark[2], \ 
	Holger Boche\footnotemark[1] \footnotemark[3] \footnotemark[5], \  
	Gitta Kutyniok\footnotemark[2] \footnotemark[4] \footnotemark[6]
}
\date{}
\maketitle

\footnotetext[2]{Department of Mathematics, Ludwig Maximilian University Munich, Germany}
\footnotetext[1]{Institute of Theoretical Information Technology, TUM School of Computation,
Information and Technology, Technical University of Munich, Germany}
\footnotetext[3]{Munich Center for Quantum Science and Technology (MCQST), Schellingstr. 4, D-80799 Munich, Germany}
\footnotetext[5]{Munich Quantum Valley (MQV), Leopoldstr. 244, D-80807 Munich, Germany}
\footnotetext[4]{Munich Center for Machine Learning (MCML), Geschwister-Scholl-Platz 1
80539 Munich, Germany}
\footnotetext[6]{Department of Physics and Technology, University of Tromsø, Norway}

\begin{abstract}
In this paper, we investigate the computational complexity of solutions to the Laplace and the diffusion equation. We show that for a certain class of initial-boundary value problems of the Laplace and the diffusion equation, the solution operator is $\# P_1/ \#P$-complete in the sense that it maps polynomial-time computable functions to the set of $\#P_1/ \#P$-complete functions. Consequently, there exists polynomial-time (Turing) computable input data such that the solution is not polynomial-time computable, unless $FP=\#P$ or $FP_1=\#P_1$. In this case, we can, in general, not simulate the solution of the Laplace or the diffusion equation on a digital computer without having a complexity blowup, i.e., the computation time for obtaining an approximation of the solution with up to a finite number of significant digits grows non-polynomially in the number of digits. This indicates that the computational complexity of the solution operator that models a physical phenomena is intrinsically high, independent of the numerical algorithm that is used to approximate a solution. 
\end{abstract}
\vspace*{1em} \textbf{Keywords} $\#P$-completeness $ \cdot $ Complexity blowup $ \cdot $ Computation complexity $ \cdot $ Diffusion equation $ \cdot $  Laplace equation $ \cdot $ Turing machine 

\section{Introduction}

For centuries, people have attempted to adequately describe physical phenomena using mathematical tools. However, it was not until the formulation of the first differential equations in the 17th century by Isaac Newton and Gottfried Wilhelm Leibniz and the resulting theory of differential equations that a milestone was reached in the mathematical description of physical phenomena. However, for practical relevance, the solution of differential equations have to be computed. Hence, algorithms have to be developed that calculate the solution efficiently or at least a sufficient approximation of the solution.

With the mathematical model of a Turing machine, the model of modern computer, introduced in 1936 by Alan Turing in his seminal paper \cite{Turi37CNAE} another milestone in automating computations was reached. It was soon postulated and believed that a numerical function is computable by a physical device if and only if it is computable by a Turing machine, known as the Physical Church--Turing Thesis. Assuming the correctness of the Physical Church--Turing thesis, a Turing machine could ideally calculate a solution of a differential equation with required precision. However, one had to realize early that Turing machines have theoretical limitations. Alan Turing already established in his previously mentioned work that the set of numbers in $\mathbb{R}$ that are computable (henceforth Turing computable) on a digital computer is only countable as there are only countably many algorithms. Similarly, there are only countably many computable real-valued functions. Subsequently, it has also been shown that a certain class of physical processes in, e.g., continuum mechanics, quantum mechanics, plasma physics, general relativity described by the wave equation 
\begin{align}
    u_{tt}(t,x)-c\Delta u(t,x)=f(t,x),
\end{align} which, under certain conditions and in the weak formulation, are well-posed in the sense of Hadamard\footnote{A problem is well-posed in the sense of Hadamard, if there exists a unique solution to the problem that continuously depends on the input data.}, cannot be simulated on a digital computer, see Section \ref{se:literature}. This is based on the fact that many mathematical operations that are used to calculate solutions to partial differential equations are in general not computable such as the Fourier transform or the differential operator \cite{Ko82NRCC}. As a result, any algorithm that requires these mathematical operations in order to calculate the desired function cannot guarantee the correctness of the output function.

In the present work, the functions we wish to compute by a Turing machine are solutions to the Laplace equation and the diffusion equation which, supplemented with inital and boundary conditions, are a well-posed class of Cauchy problems whose solutions have a closed form and satisfy many useful properties. The closed form of the solutions allows to quantify the complexity of the solutions. Under certain conditions, we show that the solutions have high complexity even though the functions from the boundary and intial conditions have low complexity. More preciseley,  we show that the solution operator maps polynomial-time computable functions to functions of the class $\# P$ or $\# P_1$, a presumably larger class of functions containing the class of polynomial-time computable functions. As a consequence, there exists no approximation, numerical or discretization operator that can compute the solution in polynomial-time for every polynomial-time computable input function unless $FP=\# P$ or $FP_1=\# P_1$ as the complexity is intrinsic to the solution operator

In order to make the previous statement more precise, we present the classes of Cauchy problems we want to discuss as well as the notions and definitions from the theory of computable analysis and complexity theory.
Let $\Omega\subset \mathbb{R}^d, d\in \mathbb{N}$, be an open and bounded domain with sufficiently smooth boundary, e.g., $\partial \Omega \in \rmC^1$. Then, we consider the Poisson equation 
\begin{align} \label{eq:poisson.eqn}
 \Delta u(x)=f(x), \quad x\in \Omega,
\end{align}  where $\Delta:=\sum_{i=1}^d \partial_{x_{ii}}$ denotes the Laplace operator, $f:\Omega\rightarrow \mathbb{R}$ an external force, and $u:\Omega \rightarrow \mathbb{R}$ is called solution of the Poisson equation. In case $f=0$, the equation \eqref{eq:poisson.eqn} is called Laplace equation. The Poisson equation, thus also the Laplace equation, are of significant importance in physics as they can describe different physical phenomena, e.g., Fick's law of diffusion, Fourier's law of heat conduction or Ohm's law of electrical conduction \cite{Evan98PDE}.  In the described phenomena, the quantity $u$ represents the chemical concentration, the temperature, or the electrostatic potential, respectively, thus describing the density of a certain  quantity. The solution of the Poisson equation describes these quantities in an equilibrium state meaning that there is no time evolution. 

The other class of Cauchy problems is given by the diffusion (or heat) equation 
\begin{align} 
	u_t(t,x)=\alpha \Delta u(t,x) \quad (t,x)\in (0,+\infty)\times \Omega,
\end{align} which is a model that describes the time evolution of the density of, e.g., the temperature, in a homogeneous and isotropic medium. Here, the function $u(t,x)$ describes the density at point $x\in \mathbb{R}^d$ and time $t>0$. The constant $\alpha>0$ is the so-called diffusion coefficient which depends on the medium and describes, e.g., in the context of the heat equation the thermal conductivity. In 1905, Albert \textsc{Einstein} showed in a seminal work that the density of Brownian particles satisfies the diffusion equation. It is well-known that the diffusion equation can be seen as a gradient flow equation, in particular as the gradient flow for the Dirichlet energy $\calE:\rmL^2(\mathbb{R}^d)\rightarrow [0,+\infty]$ defined by
\begin{align} \label{energy}
	\calE(v): =
	\begin{cases}
		\int_{\Omega}\vert \nabla v(x)\vert^2 \dd x \quad \text{if } \nabla v \in \rmL^2(\Omega),\\
		+\infty \quad \quad \quad \quad\quad \,\, \,\text{otherwise,}
	\end{cases}
\end{align} with respect to the $\rmL^2$-distance. Therefore, the $\rmL^2$-distance is the dissipation mechanism of the gradient flow, see, e.g., \textsc{Ambrosio} et al. \cite[Remark 2.3.9., p. 49]{AmGiSa05GFMS}. Denoting by $u$ a solution to the diffusion equation, from the calculation 
\begin{align*}
	\frac{\dd }{\dd t}\calE(u(t,\cdot))&= \langle -D\calE(u(t,\cdot)),u_t(t,\cdot)\rangle_{\rmL^2} =- \int_{\Omega}\vert \Delta u(t,x)\vert^2 \dd x \leq 0
\end{align*} for all $t> 0$, it follows that the energy functional \eqref{energy} decreases over time, meaning that the energy of the system dissipates over time. Therefore, the solution tends to minimize the energy over time, and the energy functional serves as a Lyapunov functional for the diffusion equation. It can also be shown that for $t\rightarrow \infty$, the solution to the heat equation converges to a solution of the Laplace equation.  Conversely, a solution to the  Laplace equation solves also the heat equation with appropriate boundary conditions. Hence, the  Laplace equation can be seen as the steady-state diffusion equation.  

As we mentioned before, the Cauchy problems to the  Laplace equation and the diffusion equation we consider in this paper are well-posed problems, i.e., for sufficiently smooth intial and boundary functions, the equations possess a unique solution and the solution depends continuously on the input data. Furthermore, there exists an explicit formula of the solution, see Evans \cite{Evan98PDE}. In Section \ref{se:Laplace} and \ref{se:diffusion}, we will see that for the explicit formula, we need to calculate Green's function which depends on the shape of the domain and is, in general, not easy to determine. However, for certain domains, Green's function can be specified. Based on this explicit formula, we will determine the complexity of solutions to the Laplace and the diffusion equation supplemented with certain initial and boundary conditions.

\subsection{Related work}\label{se:literature}
The complexity of solutions to differential equations has been studied in very few articles and for very few cases. In the following, we present results on ordinary and partial differential equations separately. 
\subsubsection{Ordinary Differential Equations}
\label{se:ODE}
In this section, we review the previous results regarding the computability and the complexity of solutions to ordinary differential equations.

The presumably first result on the computability of solutions was given by \textsc{Pour-El} and \textsc{Richards} \cite{PouRic79ACOE}. The authors showed that there exists a ordinary differential equation
\begin{align}\label{ODE}
\begin{cases}
    u'(t)=F(t,u(t)), \quad t>0\\
    u(0)=0,
\end{cases}
\end{align} such that $F$ is computable on the rectangle $[0,1]\times [-1,1]$, but no solution to \eqref{ODE} is computable on any interval $[0,\delta], \delta>0$. It has also been shown that if $F$ is computable and the ordinary differential equation \eqref{ODE} possesses a unique solution, then the solution is also computable.

\textsc{Ko} \cite{Ko83CCOD} has shown the following improved results: There exists a polynomial-time computable function $F$ such that every solution to \eqref{ODE} supplemented with the initial condition $u(0)=0$ is not computable on any interval $[0,\delta]$. The second main result says that for any recursive function $\phi$, there exists a polynomial-time computable function $F$ such that the ordinary differential equation \eqref{ODE} supplemented with the initial condition $u(0)=0$ has a unique solution on $[0,1]$, but is not computable by any oracle Turing machine operating in time with respect to $\phi$.
In his main result, \textsc{Ko} showed the following complexity result: if the function $F$ is polynomial-time computable and satisfies a weak Lipschitz condition, then the unique solution is polynomial-space computable. Furthermore, he showed that there exists a polynomial-time computable function $F$ which satisfies this weak Lipschitz condition such that the unique solution $u$ is
not polynomial time computable unless P = PSPACE.
\textsc{Kawamura} \cite{Kawa09LCOD} complemented this result by showing that, under the conditions mentioned above, the unique solution is polynomial-time computable if and only if P = PSPACE. 

\textsc{Boche} and \textsc{Pohl} \cite{BocPoh21CBSA} have shown a more concrete complexity result: there exists a polynomial-time computable and differentiable input function $x$ such that solutions to the linear ordinary differential equation given by
\begin{align*}
\begin{cases}
        y'(t)+\alpha y(t)&=\beta_0 x'(t)+\beta_1 x(t),\\
        x(0)&=x_0,\\
        y(0)&=y_0,
\end{cases} 
\end{align*} are in some sense complete for $\#P$ (see Section \ref{se:complexity_classes}), where the initial values and the coefficients are all polynomial-time computable real numbers. A similar result has been obtained by the same authors for the higher order system
\begin{align}
\begin{cases}
        \sum_{n=1}^N \alpha_n y^{(n)}(t)&=\sum_{n=1}^N \beta_n x^{(n)}(t),\\
        x^{(n)}(0)&=x_n,\\
        y^{(n)}(0)&=y_n,\quad n=1,\dots,N.
\end{cases} 
\end{align} 
Apart from these results on ODEs, \textsc{Ko} \cite{Ko92CCIE} has also investigated the computational complexity of Volterra integral equations of first and second kind given by
\begin{align*}
 f(t) = \int_0^t K(t,s, x(s))\,ds
\end{align*} and 
\begin{align*}
  x(t) = f(t) + \int_0^t K(t,s, x(s))\,ds,
\end{align*} respectively. Formally deriving the Volterra integral equation of second kind leads to an ordinary differential equation. However, a solution to this equation is not necessarily differentiable. It has been shown that under the assumption that the functions $K$ and $f$ are polynomial-time computable, $K$ satisfies a global  Lipschitz condition, the solution to the Volterra integral equations of second kind is unique and has a polynomial modulus of continuity, and the solution is bounded between $-1$ and $1$, then the solution is polynomial-space computable. A similar result has been obtained in the case $K$ satisfies a local Lipschitz condition in which case it has been shown that the solution is exponential-space computable. 

More complexity, computability, and non-computability results on ordinary differential equations can be found in the handbook \cite[Chapter 3, pp. 71]{GraZho21TACP}.

\subsubsection{Partial Differential Equations}
We next recall the previously derived key results regarding the computability and the complexity of solutions to partial differential equations.

The first known result for partial differential equations was provided by \textsc{Pour-El} and \textsc{Richards} \cite{PouRic83CDNO} who showed that solutions to the $d$-dimensional wave equation supplemented with certain initial and boundary conditions are for any $d\in \mathbb{N}$ non-computable at time $t=1$. On the other hand, it has been shown that solutions to the Laplace equation and the heat equation are computable.

The result concerning the non-computability of solutions to the wave equation has been improved subsequently by \textsc{Pour-El} and \textsc{Zhong} \cite{PouZho97WENC}, who showed that the three-dimensional wave equation, supplemented with computable and differentiable initial data, is nowhere computable. This result has been improved by \textsc{Boche} and \textsc{Pohl} \cite{BocPoh20TMCT}, who showed that the initial data can be chosen to be computable and continuously differentiable with compact support in an annulus such that the first derivative is absolute continuous.

Under an extended notion of computability for Turing machines, nameley the Type-2 theory of effectivity, which allows to consider computability questions on more general topological spaces \cite{Weih00CA},  \textsc{Weihrauch} and \textsc{Zhong} \cite{WeiZho01LSPT, WeiZho06SPTM} have shown that if initial data are functions of a Sobolev space, then the solution operator to the linear and nonlinear Schr{\"o}dinger equation are computable. Moreover, they showed that if the initial data are $L^p$-functions with $p\neq 2$, then the solution operator to the linear Schr{\"o}dinger equation are non-computable.

A result on the computational complexity of solutions to the Poisson equation has been established by \textsc{Kawamura, Steinberg}, and \textsc{Ziegler} \cite{KaStZi17CCPE}. More precisely, the authors established two main theorems on the computational complexity of solution to
\begin{align}\label{eq:poisson}
    \begin{cases}
        \Delta u = f\quad \text{in } \Omega\\
        u =  g \qquad  \text{on } \partial \Omega
    \end{cases}
\end{align} where $\Omega\subset \mathbb{R}^d$ is the unit ball. The first result demonstrates that for all smooth and polynomial-time computable functions $g:\partial \Omega\rightarrow \mathbb{R}$ with $f=0$, the unique classical solution $u: \Omega\rightarrow \mathbb{R}$ to Equation \eqref{eq:poisson} is polynomial-time computable on the closed unit ball if $FP=\#P$. In the second result, the authors demonstrated that for all smooth and polynomial-time computable functions $f: \Omega\rightarrow \mathbb{R}$ with $g=0$, the unique classical solution $u: \Omega\rightarrow \mathbb{R}$ to Equation \eqref{eq:poisson} is polynomial-time computable on the closed unit ball if and only if $FP=\#P$. 
Their hardness result for the Laplace equation has been improved by our result on the Laplace equation in the sense that we show that the solutions can be computed on the open unit ball in $\#P_1$ and this computation is optimal.
Their hardness result for the Laplace equation has been improved by our findings. Specifically, we demonstrate that the solutions can be computed on the open unit ball in $\#P_1$, and this computation is optimal. In contrast, the second result concerning the Poisson equation shows optimality in $\#P$. Similar to our approach, their methodology also relies on the ability to express the solution using Green's function, as detailed in Section \ref{se:Laplace}, and the computational complexity of integration showed in Theorem \ref{Friedman1}. However, the approximation techniques they employ are completely different. While we achieve an approximation of the solution by truncating its expansion in terms of spherical harmonics, they construct an polynomial-time approximation of the solution by truncating the unbounded Poisson kernel near its singularity.

Recently, the authors \textsc{Koswara}, \textsc{Pogudin}, \textsc{Selivanova}, and \textsc{Ziegler} in \cite{KPSZ21BCSS} investigated the bit-complexity of solutions and discretization schemes for the linear partial differential equation:
\begin{align} \label{eq:linEQ}
    \begin{cases}
    u_t(t,x) = \mathcal{A}u (t,x) \quad \, \, \,(t,x)\in (0,1)\times \Omega,\\
    u(0,x)=\varphi(x), \qquad \quad x\in \Omega,\\
    \mathcal{L}u(t,x)\vert_{(0,1)\times \partial \Omega}=0\qquad \quad \quad \, (t,x)\in (0,1)\times \partial \Omega,
    \end{cases}
\end{align} where $\Omega=[0,1]^d$ denotes the unit cube in $\mathbb{R}^d$, and $\mathcal{A}$ and $\mathcal{L}$ are linear differential operators, with $\mathcal{A}$ having a higher order than $\mathcal{L}$. Specifically, they demonstrated the following under certain regularity conditions of the solution:

\begin{itemize}
\item The solution to \eqref{eq:linEQ} can be computed in PSPACE.
\item For $\mathcal{A}=\sum_{j=1}^d B_j\partial_{x_j}$, the solution \eqref{eq:linEQ} can be computed in $\#P$ assuming that the matrices $B_j$ mutually comute.
\item For all polynomial-time computable initial conditions $\varphi$, the solution to the heat equation is not polynomial-time computable unless $FP_1=\#P_1$.
\end{itemize}  Furthermore, they demonstrated that if a particular difference scheme for \eqref{eq:linEQ} converges under certain conditions to the solution, then that solution is computable in $\#P$.

Similarly, our results further refine the findings on the heat equation, highlighting the optimality of the class $\#P_1$.

\subsection{Contributions}
In this work, we investigate the computational complexity of solutions to the Laplace and the diffusion equation for various initial and boundary conditions. More precisely, we show that under certain conditions, the solution operator maps polynomial-time computable intial and boundary functions to the computable unique solution of the Laplace equation and the diffusion equation, and the solution is polynomial-time computable if (and only if) $FP=\#P$ and $FP_1\#P_1$, respectively. Unless $FP\neq \#P$ and $FP_1\neq \#P_1$ (see Section \ref{se:complexity_classes}), this implies that the computation of the solution operator maps low complexity functions to high complexity functions meaning that the computation time for obtaining an approximation of the solution with up to $n\in \mathbb{N}$ significant digits grows non-polynomially in $n$. Hence, the solution operator has intrinsically high complexity in the sense that for fixed polynomial-time initial/boundary data it outputs. As a consequence, we show that there does not exist a numerical approximation scheme that can compute the solution in polynomial-time for all polynomial-time computable data. 
We want to stress that our complexity result is understood in an non-uniform way by fixing the initial or boundary condition, see \cite{KawCoo10CTOA} for operator complexity. However, we believe that the results can be generalized as operator complexity as it was mentioned in \cite{KawCoo10CTOA} that most of the work is done by showing the pointwise complexity result. 
We wish to emphasize that our complexity result is understood in a non-uniform way when the initial or boundary condition is fixed, as detailed in \cite{KawCoo10CTOA} concerning operator complexity. Nevertheless, we believe that our results can be generalized to operator complexity as \cite{KawCoo10CTOA} mentioned that the bulk of the proof involves demonstrating the pointwise complexity result.

As it is well-known that $N\neq NP$ implies $FP\neq \#P$, our result also connects the notoriously hard P vs. NP problem from structural complexity theory to the complexity of physical phenomena.

In the following we summarize our contributions:
\begin{itemize}
    \item The Laplace and the diffusion equation supplemented with polynomial-time computable initial and boundary conditions, have solutions that can be computed in $\# P$-complete, meaning that the computation of the solution in $\# P$ is essentially optimal.
    \item The solution operator that maps the intial and boundary functions to the solution of the Laplace and the diffusion equation have intrinsically high complexity. 
    \item There does not exist numerical algorithm that can compute the solutions in polynomial-time for all initial and boundary functions, unless $FP= \#P$.
    \item Physical phenomena that can be described by the Laplace and diffusion equation have high complexity.  
    \item We provide a general approach of determining the computational complexity of functions, in particular solutions to partial differential equations that have a closed form.
\end{itemize}

Investigations of the complexity of solutions to partial differential equations are to the authors' best knowledge not existent in the literature.

\subsection{Outline}
The remainder of the paper is organized as follows: In Section \ref{se:comp.analysis}, we introduce notions from computable analysis and complexity theory. While Section \ref{se:Laplace} is devoted to the complexity results for the Laplace equation on the sphere in any dimension, in Section \ref{se:diffusion}, we show complexity results for the one dimensional diffusion equation on different domains in space: a compact interval and the positive real line. In Section \ref{se:conclusion}, we finish with a conclusion and an outlook. In Section \ref{se:appendix}, some mathematical tools we use throughout the paper are provided.  

\section{Computable analysis and complexity} \label{se:comp.analysis}

In this section, we introduce notions and results from the theory of computable analysis as well as complexity theory that are relevant for our analysis. This section mainly relies on the definitions and notions of \textsc{Pour-El} and \textsc{Richards} \cite{PouRic89CIAP}, \textsc{Friedman} and \textsc{Ko} \cite{KoFri82CCRF,Frie84CCMI} as well as \textsc{Arora} and \textsc{Barak} \cite{AroBar09CCMA}. For a detailed treatise of the theory of computable analysis, we refer the interested reader to \cite{AroBar09CCMA} and \textsc{Weihrauch} \cite{Weih00CA}.

\subsection{Computable Analysis}
\subsubsection{Computation by a Turing machine}

In this section, we introduce the notion of computability, more specifically the notion of a computable number and a computable function. Since there are different notions of computabiliy, e.g., Banach-Mazur computability, Turing (also Borel--Turing) computability, weak computability, $L^p$-computability, Type-2 Turing computability etc.,  we specifically rely on the notion of Turing computability for continuous functions in the sense of Turing \cite{Turi37CNAE}. Equivalent definitions have been provided by other authors, see e.g., \textsc{Grzegorczyk} \cite{Grze57CRCF}, and \textsc{Pour-El} and \textsc{Caldwell} \cite{PouCal75SDCF}. We refer the interested reader to \textsc{Avigad} and \textsc{Brattka} \cite{AviBra14TLTI} for a historical view of the notion of computability.

The Turing machine is the theoretical concept on which every modern computer is based on. In contrast to a digital computer, it has no physical limitations in theory. On the other hand, every Turing machine defines the theoretical limits of the digital computer.
\begin{defi}(Computable number) \label{def:TM}
    A number $t\in \mathbb{R}$ is said to be \emph{computable}, if there exists a Turing machine TM with input $n \in \mathbb{N} $ and output $\gamma(n)=TM(n)\in \mathbb{Q} $, such that
\begin{align}\label{computable.number}
    \vert t-\gamma(n)\vert\leq 2^{-n}, \quad \text{for all }n\in \mathbb{N}.
\end{align}
In this case, we say that $\gamma(n)$ \emph{binary converges to} $t$, and we write
$\mathbb{R}_c \subsetneq \mathbb{R}$ for the set of all computable real numbers.
\end{defi}
The Turing machine TM defines the limit of real numbers that are computable in the sense of Definition \ref{def:TM}.  However, it does not specify the number of iterations (i.e. computation time) that are required for the Turing machine to calculate $\gamma(n)$ for a given input $n\in \mathbb{N}$. Since we are interested  in determining the computational complexity of certain problems, the next definition quantifies the number of  iterations that are required to approximate $t\in \mathbb{R}_c$ as $n$ increases. 
\begin{defi}(Polynomial-time computable number)
    Let $t\in \mathbb{R}_c$ be a computable number. We say that the \emph{computational complexity of $t$ is bounded by a function} $q:\mathbb{N}\rightarrow \mathbb{N}$, if there exists a Turing machine TM such that the function $\gamma$ computed by the Turing machine satisfies \eqref{computable.number} and such that, on input $n$, the Turing machine stops after at most $q(n)$ steps. The number $t\in \mathbb{R}_c$ is said to be \emph{polynomial-time computable}, if its computational complexity is bounded by a polynomial $q$.
\end{defi}
In order to define computable functions, we employ the concept of a function-oracle Turing machine. An oracle Turing machine is a Turing machine that is equipped with an function-oracle which calculates the value of $\gamma$ from Definiton \ref{def:TM} in a single step. However, the previous definition suggests that the calculation of $\gamma$ might take some computational time. This concept allows to quantify the computational complexity of a function distinct from the computational complexity of the input data, which requires a Turing machine itself to be approximated. 

\begin{defi}(Computable function) \label{def:comp.func}
    Let $x:[a,b]\rightarrow \mathbb{R}$ be a real function. Then $x$ is said to be \emph{computable on the interval}  $[a, b] \subset \mathbb{R}$, if there exists a function-oracle Turing machine TM, such that for each $t \in [a, b]$ and each $\gamma$ that binary converges to $t$, the
function $\tilde{x}(n) = TM_\gamma (n)$ computed by TM with oracle $\gamma$ binary converges to $x(t)$, i.e., if
\begin{align}\label{computable.function}
    \vert x(t)-\tilde{x}(n)\vert\leq 2^{-n}, \quad \text{for all }n\in \mathbb{N}.
\end{align}
\end{defi}
 Henceforth, we refer to a function-oracle Turing machine simply as Turing machine for reasons that will be elaborated in the next section. The following result shows that any computable function is continuous on its domain.
\begin{pro}
Let $x : [a, b] \rightarrow \mathbb{R}$ be a computable function on $[a, b]$, then $x \in \rmC([a, b])$.
\end{pro}
Similar to the definition of a polynomial-time computable number, we can provide a definition of a polynomial-time computable function. However, here we allow only oracle functions $\gamma:\mathbb{N}\rightarrow D_1$ with values in the set of dyadic rationals $D_1$, i.e., rational numbers in the interval $[0,1]$ of the form
\begin{align}
    t_{j,n} = j \cdot 2^{-n}, \quad 0 \leq j \leq 2^n,
\end{align}
for some $n \in N$.
\begin{defi}(Polynomial-time computable function)\label{def:ptc.oracle}
   Let $x : [a, b] \rightarrow \mathbb{R}$ be a computable function. We say that the \emph{complexity of $x$ is bounded by a function} $q : \mathbb{N}\rightarrow \mathbb{N}$, if there exists a function–oracle Turing machine TM, which computes $x$ such that for all $\gamma:\mathbb{N}\rightarrow D_1$, that binary converge to a real number $t \in [a, b]$ in a way that the denominator of $\gamma$ grows linearly in $n\in \mathbb{N}$, and
for all $n\in \mathbb{N},$ there holds
\begin{align}\label{computable.function.poly}
    \vert x(t)-TM_\gamma(n)\vert\leq 2^{-n}
\end{align}
after a computation time of at most $q(n)$. The function $x:[0, 1] \rightarrow \mathbb{R}$ is said to be \emph{polynomial-time computable}, if its complexity is bounded by a polynomial $q$.
\end{defi}

\subsubsection{Computation on a dyadic grid}
The definition of the computability of a function $f$ by a Turing machine relies on an oracle which, for inputs $t\in [0,1]$ and $n\in \mathbb{N}$, computes an output $\tilde{t}\in  [0,1]\cap \mathbb{Q}$ such that $\vert t-\tilde{t}\vert<2^{-n}$. This input-output mechanism is done by the oracle in one step. Hence by approximating the number $f(t)$, the complexity of approximating the number $t$ is not taken into account, thus focusing purely on the complexity of $f$. 
In this paper, we want to introduce an equivalent definition of computability which does not rely on function-oracle Turing machines. In this
approach, we restrict the domain of $f$ to the discrete set $D_1\subset [0, 1] \cap \mathbb{R}_c$, where $D_1$ denotes again the set of dyadic rationals. The definition of polynomial-time computability will be phrased in Definition \ref{def:ptc.dyadic}.

With the following definition, we first equivalently describe any dyadic rational by a binary string, i.e., a finite sequence of zeros and ones.
\begin{defi}(Dyadic rational)
    A \emph{dyadic rational} consists of the symbol $'+'$ or $'-'$  followed by a (possibly empty) string of 0’s and 1’s which starts (if it is nonempty) with 1, followed by a decimal point, followed by a second (nonempty) string of 0’s and 1’s. The set of all dyadic rationals is denoted by $D$.
\end{defi} 
It is clear that $D_1\subset D$ and $D_1$ consists of all $d \in D$ which begin with $'+'$ and for which the string to the left of the decimal point is empty. With $\mathrm{tnd}(d)$ we denote the total number of digits in $d$, and with $\mathrm{pcs}(d)$ the total number of digits to the right of the decimal point in $d$. 
Furthermore, for any $t \in \mathbb{R}$, we write $d \sim t$, if $\vert d -t\vert \leq 2^{-\mathrm{pcs}(d)}$. This describes how well the number $t$ is approximated by $d$. 

Next, we define the dyadic analog to the Turing machine that can approximate a certain function in this model.
\begin{defi}(Dyadic approximation function)
A \emph{dyadic approximation function} is a pair $(g, i)$ with $i \in \mathbb{N}$ and with a function $g : D_1 \rightarrow D$, so that, for all $n \in \mathbb{N}$, there exists $m \in \mathbb{N}$ such that for every $d \in D_1$ with $\mathrm{tnd}(d) \geq m$, we have $\mathrm{tnd}(g(d)) \geq n$.
\end{defi}
Now, a function $f : [0, 1] \rightarrow \mathbb{R}$ is said to be \emph{approximable by a dyadic approximation function} $(g, i)$, if for all $t \in [0, 1]$ and all $d \in D_1$ with $tnd(d) \geq i$ and $d \sim t$, we have  $g(d) \sim f(t)$. In this case, we write $(g, i) \sim f$.
It has been proven by Friedman \cite{Frie84CCMI} that if $(g,i)$ is a dyadic approximation that approximates some $f$, then $f$ is uniformly continuous and uniquely determined by $g$.

\begin{defi}(Modulus)
Let $h:\mathbb{N} \rightarrow \mathbb{N}$ be a function on $\mathbb{N}$ and let $(g, i)$ be a dyadic approximation function. We call $h$ a \emph{modulus} for $(g, i)$, if, for every $n \in \mathbb{N}$ and all $d \in D_1$ with $tnd(d) \geq h(n)$, we have $tnd(g(d)) \geq n$.
\end{defi}  
It has also been shown in \cite{Frie84CCMI} that if $h$ is a modulus for a dyadic approximation $(g,i)$ and $(g,i)$ approximates some function $f:[0,1]\rightarrow \mathbb{R}$, then there exists some constant $c\in \mathbb{N}$ such that 
\begin{align}
 \vert x-y\vert <2^{-h(n+c)+c}
\end{align} implies 
\begin{align}
    \vert f(x)-f(y)\vert <2^{-n} \quad \text{for all } x,y\in [0,1], n\in \mathbb{N}.
\end{align}
Thus, a modulus is related to the modulus of continuity which quantitatively measures the uniform continuity of a function. In fact, one can define computability in terms of the modulus of continuity, see \cite{KaStZi17CCPE}.

Similarly, we say that a function $g : D_1 \rightarrow D$ is polynomial-time computable, if there exists a polynomial $q$ and a Turing machine $TM:D_1\rightarrow D$ such that the Turing machine computes $g$ and such that the computation time is at most $q(tnd(d))$ for every $d \in D_1$. 
\begin{defi}(Polynomial-time computable on dyadic grids) \label{def:ptc.dyadic}
A function $f : [0, 1] \rightarrow \mathbb{R}$ is said to be \emph{polynomial-time computable} on dyadic grids, if there exists a polynomial-time computable function $g : D_1 \rightarrow D$ and an $i \in N$ so that $(g, i)$ has a polynomial modulus, and $(g, i) \sim f$.
\end{defi}

We wish to remark that in this definition of polynomial-time computability, the computation time for approximating $t \in [0, 1]$ by a dyadic rational $d \in D_1$ is now taken into account and does not rely on a function-oracle anymore. Now, one might ask whether there does exist a relation between these seemingly different notions of polynomial-time computability. In fact, Friedman showed in \cite{Frie84CCMI} that both definitions are equivalent. 
\begin{theorem}(Friedman)
Let $f : [0, 1] \rightarrow \mathbb{R}$ be a given
function. Then $f$ is polynomial-time computable in the sense of Definition \ref{def:ptc.oracle} if and only if $f$ is polynomial-computable in the sense of Definition \ref{def:ptc.dyadic}.
\end{theorem}

In regard of the multi-dimensionality of the solution to the partial differential equations we consider in this paper, it is necessary to extend the notion of computability to multivariable function. This can be done in the following natural way: \emph{An $m$-dimensional dyadic approximation function} is a pair $(g,i)$ with $i\in \mathbb{N}$ and a function $g:D_1^m\rightarrow D$, so that, for all $n\in \mathbb{N}$ and for some function $\alpha$, $tnd(d_1), \dots, tnd(d_m)\geq \alpha(n)$ implies $tnd(g(d_1,\dots,d_m))\geq n$. Such a function  $\alpha:\mathbb{N}\rightarrow \mathbb{N}$ is called a \emph{modulus} of $g$. Let $f:[0,1]^m\rightarrow \mathbb{R}$. We write $(g, i) \sim f$ if for all $d_1 \sim x_1, \dots, d_m\sim x_m,$, we have $g(d_1,\dots,d_m)\sim f(x_1,\dots,x_m)$. We say that $f:[0,1]^m\rightarrow \mathbb{R}$ is \emph{polynomial-time computable} if there is a $(g, i) \sim f$ such that $g$ is polynomial-time computable and $(g,i)$ has a polynomial modulus.

This definition of computability can naturally be extended to any rectangular domain \( \Omega \) in \( \mathbb{R}^d \). In the case where \( \Omega \) is the unit ball or the sphere, one can convert to spherical coordinates to obtain a rectangular domain. For unbounded domains \( \Omega \), we define a function as computable if it is computable on any compact rectangular subset contained within \( \Omega \). Similarly, a function is termed polynomial-time computable if it can be computed in polynomial time on any compact rectangular subset of \( \Omega \).

\subsection{Complexity classes} \label{se:complexity_classes}
In this section, we want to introduce some complexity classes that capture and define the complexity of evaluating a given function in a reasonable way. In particular, we introduce complexity classes for decision, counting, and function problems. Then, with the definition of the computabilty of a function on a dyadic grid, we will be able to formulate the problem of evaluating a function as a counting problem. \\\\
\textit{Decision problems and the classes P and NP:}
In order to investigate the computational complexity of solutions to partial differential equations, we need to introduce appropriate complexity classes. The best known complexity classes are $P$ and $NP$ which consist of \textit{decision problems}. Decision problems are problems that require a  `yes' or `no' answer. The class $P$ consists of all decision problems that are solvable in polynomial-time by a deterministic Turing machine meaning that the computational complexity grows polynomially in the input size. The class $NP$ consists of all decision problems for which a given answer can be verified in polynomial-time by a deterministic Turing machine. It is obvious that $P\subset NP$. However, the question whether $P= NP$ or $P\subseteq NP$ remains a major open problem in structural complexity theory and belongs to the famous Millennium Prize Problems . \\\\
\textit{Counting problems and the classes $\#P$ and $\#P_1$:} Another complexity class is given by the set of \textit{counting problems}, which does not ask whether a given problem in $NP$ has a solution but enumerates the number of solutions. For a more formal definition of $\#P$, let $\lbrace 0,1\rbrace^n$ be the set of all words
of length $n\in \mathbb{N}$ in the alphabet $\Sigma=\lbrace 0,1\rbrace$ consisting of 0 and 1, and let $\Sigma^*$ denote the set of all finite words in the alphabet $\Sigma$. For a given string $x\in \Sigma^*$, we denote the length of $x$ with $len[x]$. Then, a \emph{function $f : \Sigma^* \rightarrow \mathbb{N}$ is in $\#P$}, if there exists a polynomial $p:\mathbb{N}\rightarrow \mathbb{N}$ and a polynomial-time Turing machine $M$, so that for every string $x\in \Sigma^*$, we have 
    \begin{align}
        f(x)=\left \vert \left \lbrace  y\in \Sigma^{p(len[x])}: M(x,y)=1 = \text{`Yes'} \right \rbrace \right \vert.
    \end{align}
Here $f(x)$ denotes the number of accepting paths (or certificates) for the input $x$. Similarly, we define the subclass of counting problems by restricting the set of counting problems to functions of the form $f:\lbrace 0\rbrace^*\rightarrow \mathbb{N}$,  where $\lbrace 0\rbrace^* =\lbrace \lbrace 0\rbrace, \{0, 0\}, \{0, 0, 0\}, \dots \}$. We denote this class with $\#P_1$. More formally, a function $f : \lbrace 0\rbrace^* \rightarrow \mathbb{N}$ is in $\#P_1$, if there exists a polynomial $p:\mathbb{N}\rightarrow \mathbb{N}$ and a polynomial-time Turing machine $M$, so that for every string $x\in \lbrace 0\rbrace^*$, there holds 
\begin{align}
        f(x)=\left \vert \left \lbrace  y\in \lbrace 0\rbrace^{p(len[x])}: M(x,y)=1 \right \rbrace \right \vert.
\end{align} 
 A very prominent and important problem that belongs to $\#P$ is the task of calculating the permanent of a matrix with entries consisting of $0$'s and $1$'s which is related to the Boson sampling problem in quantum computing \cite{AarArk13CCLO}. In fact, this problem is even $\#P$-complete meaning that any problem in the complexity class $\#P$ can be reduced in polynomial-time to the $\#P$-complete problem. Hence, if there exists a polynomial-time Turing machine that solves a $\#P$-complete problem, then any other problem in the same complexity class can be solved in polynomial-time. In other words, the $\#P$-complete problems are the hardest among all problems in $\#P$. Similarly, one can define the property of being complete for other complexity classes. Another problem in $\#P$ is the graph coloring problem which asks for the number of admissible colorings using $k\in  \mathbb{N}$ colors for a particular graph.  \\\\
\textit{Function problems and the classes $FP$ and $FP_1$:}
Similar to the class of decision problems that can be solved in polynomial-time, we can define the class of counting problems that can be solved by a function-oracle Turing machine in polynomial-time denoted by $FP$ and $FP_1$ which are also called \textit{function problems}. Formally, the classes are defined as follows:
\begin{itemize}
    \item A function $f : \lbrace 0,1\rbrace^* \rightarrow \mathbb{N}$ belongs to $FP$, if it can be computed by a deterministic Turing machine in polynomial-time.
     \item A function $f : \lbrace 0\rbrace^* \rightarrow \mathbb{N}$ belongs to $FP_1$, if it can be computed by a deterministic Turing machine in polynomial-time.
\end{itemize}
By the definition of the classes $FP$ and $FP_1$, it is evident that $FP\subset \#P$ and $FP_1\subset \#P_1$. Similar to the $P$ vs. $NP$ problem, the question arises whether $FP= \#P$ and $FP_1= \#P_1$. Furthermore, it can be shown that $FP= \#P$ implies $P=NP$. As the equality $P=NP$ would have immense consequences in many fields, it is widely assumed that $FP \subsetneq \#P$ and $FP_1 \subsetneq \#P_1$.
It is easy to see that we can identify the set of finite strings $\Sigma^*$ over $\Sigma$ with the set of dyadic rationals $D_1\cap[0,1]$ in a canonical way: to a sequence $d_1d_2d_3\dots d_n$, we uniquely associate the binary number $d=0. d_1d_2d_3\dots d_n \in D_1\cap[0,1]$. This allows to relate polynomial-time computable functions with the complexity classes $\#P$ and $\#P_1$ as well as $FP$ and $FP_1$. 

\subsection{Computational complexity of integration}
Next, we present some essential characterizations of the complexity classes $\#P_1$ and $\#P$ in terms of integration due to Friedman \cite{Frie84CCMI}. These results allow to determine the computational complexity of solutions to the Laplace and the heat equation as the exact solution can be expressed in terms of integrals, see Section \ref{se:Laplace} and \ref{se:diffusion}.

\begin{theorem} \label{Friedman1} 
The following statements hold true:
\begin{itemize}
\item[1)] For all polynomial-time computable functions $g:[0,1] \rightarrow \mathbb{R}$, the function $f(x)=\int_0^x g(y)\mathrm{d}y$ is polynomial-time computable if and only if $FP=\# P$.
\item[2)] For all polynomial-time computable functions $g:[0,1]\rightarrow \mathbb{R}$, the number $\int_0^1g(y)\mathrm{d}y$ is polynomial-time computable if and only if $FP_1=\# P_1$.
\end{itemize}
\end{theorem}




Hence, if the widely assumed statement $FP \subsetneq \#P$ holds true, there exists a polynomial-time computable function $g$ such that the function $x\mapsto f(x)=\int_0^xg(y)\mathrm{d}y$ is not polynomial-time computable.
Similarly, if $FP_1 \subsetneq \#P_1$ holds true, there exists a polynomial-time computable function $g$ such that the value $\int_0^1g(y)\mathrm{d}y$ is not polynomial-time computable. However, \textsc{Müller} \cite{Mull87UCCT} has shown that if $h$ is analytical and polynomial-time computable, then the integral function is also polynomial-time computable independent of the class $FP$ being equal to $\#P$ or not. The idea for the proof is that every analytic function can be expanded by a power series which allows to easily calculate the integral function by calculating its coefficients. But, in general, the integral function of a polynomial-time computable function, is not polynomial-time computable if $FP\subsetneq \#P$.

With characterizations of the classes  $\#P_1$ and $\#P$ in terms of integrals, we are able to show that solutions to certain Cauchy problems for the Laplace equation and the diffusion equation can be computed in the classes $\#P_1$ and $\#P$, respectively. In fact, in Theorem \ref{th:Laplace.2d}, \ref{th:Laplace.d}, \ref{th:diffusion.compact}, we show that the solutions are even $\#P_1$-complete and in Theorem \ref{th:diffusion.neumann} that they are $\#P$-complete meaning that the computation of the solutions in $\#P_1$ and $\#P$, respectively, are essentially optimal. The proofs of the above mentioned theorems have all a similar structure:
\begin{itemize}
    \item[1.] Showing that the solution to the partial differential equation is Turing computable by constructing a Turing machine that satisfies \eqref{computable.function}.
     \item[2.] Giving a lower bound for the computational time complexity of solutions for arbitrary but fixed input data.
     \item[3.] Proving an upper bound for the computational complexity of solutions by constructing a specific input data. This proves the completeness of the solution with respect to the complexity class.
\end{itemize}

\section{The Laplace equation}\label{se:Laplace}
For the following analysis, we assume that $D:=\lbrace x\in \mathbb{R}^d: \vert x\vert_2\leq 1\rbrace, d\in \mathbb{N}$ with $\vert \cdot\vert_2$ being the Euclidean distance in $\mathbb{R}^d$. Further, we denote by $S^{d-1}=\partial D$ the $d$-sphere. Then, we supplement the Laplace equation with non-homogeneous Dirichlet boundary conditions
\begin{align}
	\begin{cases}\label{laplace}
		\Delta u=0 \quad &\text{for } x\in D,\\
		u(x)=g(x)\quad  &\text{for }x\in  \partial D,
	\end{cases}  
\end{align}  where $g\in \rmC(S^{d-1})$. It can be shown (see Evans \cite[Theorem 15, p. 41]{Evan98PDE}) that, if $g\in \rmC(S^2)$, the function $u$ defined by
\begin{align}\label{sol.laplace}
	u(x)=-\int_{S^{n-1}} \frac{\dd G}{\dd \nu }(x,y)\, g(y)\, \mathrm{d}S(y) ,\quad x \in D,
\end{align} solves \eqref{laplace} with 
\begin{align*}
	G(x,y)=\frac{1-\vert x\vert_2^2}{d V_d\vert x-y\vert_2^d}, \quad x,y\in D
\end{align*} being the so-called Green's function where $V_d$ denotes the volume of the unit ball in $\mathbb{R}^d$. In particular, it has been shown that $u$ is smooth and even analytic in $D$. For different boundary conditions and different shapes of the domain $\Omega$, Green's function looks in general differently and can also often not be represented explicitly.

Since our proof of the complexity result requires slightly different methods for $d=2$ and the general case $d\geq 3$, we consider these cases separately. 
\subsection{The case d=2}
For our main result, we need to express the solution \eqref{sol.laplace} in polar coordinates, i.e., we consider 
\begin{align}
	u(r,\vartheta)= \frac{1}{2\pi}\int_{-\pi}^{\pi} g(\tau) \frac{1-r^2}{1-2r\cos(\vartheta-\tau)+r^2}  \mathrm{d}\tau.
\end{align}

Now, we state our main theorem:

\begin{theorem} \label{th:Laplace.2d}
Let $r_0 \in \mathbb{R}_c\cap [0,1)$ be a polynomial-time computable number. We denote by $S$ the solution operator that maps $g\in \rmC(S^1)$ to the unique solution $u=Sg$ of the boundary value problem of the Laplace equation \eqref{laplace}. Then, for all polynomial-time computable functions $g\in \rmC(\partial D)$, the mapping $[0,r_0]\times[0,2\pi] \ni (r,\vartheta) \mapsto u(r,\vartheta)=(Sg)(r,\vartheta)$ is computable, and it is polynomial-time computable if and only if $FP_1=\# P_1$.
\end{theorem}
\begin{proof}
\textbf{Ad Computability:} Since $g$ is periodic and continuous, we can expand $g$ as a Fourier series by
	\begin{align*}
    g(\theta)=\frac{a_0}{2}+\sum_{k=1}^\infty \left( a_k \sin(k \theta)+b_k \cos (\theta)\right), \quad \theta\in [0,2\pi],
	\end{align*} where  $a_0=\frac{1}{\pi}\int_{0}^{2\pi} g(\tau)\mathrm{d}\tau$ as well as
\begin{align*}
 a_n=\frac{1}{\pi}\int_{0}^{2\pi}  g(\tau)\sin(n\tau)\mathrm{d}\tau,\quad b_n=\frac{1}{\pi}\int_{0}^{2\pi}  g(\tau)\cos(n\tau)\mathrm{d}\tau, \quad n\in \mathbb{N}
\end{align*} are the Fourier coefficients. We note that the series is uniformly convergent. Then, it can be shown that the solution to the boundary value problem is given by
\begin{align*}
u(r,\theta)=\frac{a_0}{2}+\sum_{k=1}^\infty  r^k  \left(a_k \sin(k \theta)+b_k \cos (k\theta)\right), \quad (r,\theta)\in [0,1]\times [0,2\pi],
\end{align*}  see \cite{EftFry14SHPD}.
We define 
\begin{align} 	\label{eq:partial.sum.laplace}
u_N(r,\theta):=\frac{a_0}{2}+\sum_{k=1}^{N K}  r^k  \left(a_k \sin(k \theta)+b_k \cos (k\theta)\right), \quad (r,\theta)\in [0,1]\times [0,2\pi],
\end{align} where $N,K\in \mathbb{N}$ and $K$ has to be determined. Then, since $r_0 \in (0,1)$, we obtain
\begin{align*}
    	\vert u(r_0,\vartheta)-u_N(r_0,\vartheta)\vert &\leq \vert \sum_{k=K N+1}^\infty  r_0^k  \left(a_k \sin(k \theta)+b_k \cos (k\theta)\right)\vert \\
	&\leq \sum_{k=K N+1}^\infty \left( r_0^k  \vert a_k \sin(k \theta)+b_k \cos (k\theta)\vert\right) \\
	&\leq \sup_{k\in \mathbb{N}}(\vert a_k\vert +\vert b_k\vert)\cdot \sum_{k=K N+1}^\infty r_0^k \\
 &\leq 4\Vert g\Vert_{\infty} \cdot \sum_{k=K N+1}^\infty r_0^k \\
  &\leq 4\Vert g\Vert_{\infty} r_0^{K N+1} \cdot \sum_{k=0}^\infty r_0^k \\
	&\leq\frac{4\Vert g\Vert_{\infty}}{1-r_0}r_0^{K N},
\end{align*} 
 where we used the fact that $\vert a_n\vert,\vert b_n\vert\leq 2 \Vert g\Vert_{\infty}$ for all $n\in \mathbb{N}$ and the convergence of the geometric series for $r_0<1$. Now, define the constant $C>0$ by 
 	\begin{align*}
C:=\frac{4\Vert g\Vert_{\infty}}{1-r_0}>0.
	\end{align*}  
 If $C\leq 1$, then we choose $K\in \mathbb{N}$ such that 
\begin{align*}
	r_0^K<\frac{1}{2},
\end{align*} and we obtain for \eqref{eq:estimate.C}, the estimate 
	\begin{align}\label{eq:estimate.C}
	\vert u(r,\vartheta)-u_N(r,\vartheta)\vert \leq C(r_0^K)^N< \frac{1}{2^N}
\end{align}  for all $(r,\vartheta)\in [0,r_0]\times[0,2\pi]$.
	 If on the other hand $C> 1$ (thus $C^{-1}<1$), then we choose $K\in \mathbb{N}$ such that 
	\begin{align*}
	r_0^K<C^{-1}\frac{1}{2}.
	\end{align*} In this case, we obtain
	\begin{align*}
		\vert u(r,\vartheta)-u_N(r,\vartheta)\vert\leq C(r_0^K)^N<C^{-(N-1)} \frac{1}{2^N}< \frac{1}{2^N} \quad \text{for all }(r,\vartheta)\in [0,r_0]\times[0,2\pi].
	\end{align*} 
\textbf{Ad Upper bound:} Since the calculation of $K\cdot N$ Fourier coefficients are each in $\#P_1$, the preceding calculations show that $FP_1^{\#P_1}$ is an upper bound for the computational complexity of $u_N$. Thus, if $FP_1={\#P_1}$, then $FP_1^{\#P_1}=FP_1$ and for every polynomial-time computable function $g$, the solution is polynomial-time computable.\\
\newline 
\textbf{Ad Completeness:} Now, we want to show that $\#P_1$ is also a lower bound for the computational complexity of the solution. 
To do so, let $\vartheta_0\in \mathbb{R}_c\cap [-\pi ,\pi]$ be any polynomial-time computable value. Then, choose the boundary condition $g:S^2\rightarrow \mathbb{R}$ by 
\begin{align*}
 g(\tau):=\tilde{g}(\tau)\frac{1-2r_0\cos(\vartheta_0-\tau)+r_0^2}{1-r_0^2}, \quad \tau \in [0,2\pi], 
\end{align*} where
\begin{align*}
\tilde{g}(\tau)=
    \begin{cases}
        h(\tau) \quad \text{for }\tau\in [0,1],\\
        h(1)+\frac{h(1)-h(0)}{1-2\pi}(\tau-1)\quad \text{for }t\in (1,2\pi],\\
    \end{cases}
\end{align*} and $h$ is chosen to be polynomial-time computable such that the calculation of 
\begin{align}
\int_{0}^{1 }h(\tau) \mathrm{d}\tau
\end{align}
is $\# P_1$-complete. Since the function $\tilde{g}$ on $(1,2\pi]$ is a linear function, it is polynomial-time computable on $(1,2\pi]$. This implies that the calculaton of $\int_{0}^{2\pi} \tilde{g}(\tau) \mathrm{d}\tau$ is, as a sum of a polynomial-time computable function and a function that is $\# P_1$-complete, also $\# P_1$-complete.  
Then, the solution of the boundary value problem \eqref{laplace} for this given $g$, is given by 
	\begin{align*}
		u(r,\vartheta)= \frac{1}{2\pi}\int_{-\pi}^{\pi} \tilde{g}(\theta)\frac{(1-2r_0\cos(\vartheta_0-\tau)+r_0^2)(1-r^2)}{(1-r_0^2)(1-2r\cos(\vartheta-\tau)+r^2)}  \mathrm{d}\tau, \quad (r,\vartheta)\in [0,1]\times [0,2\pi].
	\end{align*}   Now,  for $(r,\vartheta)=(r_0,\vartheta_0)$, we obtain
	\begin{align*}
		u(r_0,\vartheta_0) &= \frac{1}{2\pi}\int_{-\pi}^{\pi} \tilde{g}(\theta)\frac{(1-2r_0\cos(\vartheta_0-\tau)+r_0^2)(1-r_0^2)}{(1-r_0^2)(1-2r_0\cos(\vartheta_0-\tau)+r_0^2)}  \mathrm{d}\tau\\
  & = \frac{1}{2\pi}\int_{-\pi}^{\pi} \tilde{g}(\theta) \mathrm{d}\tau.
	\end{align*} Since $ \frac{1}{2\pi}$ is polynomial-time computable, the computation of $u(r_0,\vartheta_0)$ is hence $\# P_1$-complete. 
\end{proof}

\subsection{The general case $d\geq 3$}
In this section, we extend the previous results to the $d$-dimensional case with $d\geq 3$. First, we wish to remark that the solution in Cartesian coordinates has the form 
\begin{align*}
	u(x)=\int_{\partial D}\frac{1-\vert x\vert_2^2}{4\pi\vert x-y\vert_2^d} g(y)\dd y, \quad x\in D.
\end{align*} In spherical coordinates, the solution is given by
\begin{align*}
	u(r,\mathbold{\vartheta},\varphi)=\int_{0}^{2\pi} \int_{[0,\pi]^{d-2}} \frac{1-r^2}{4\pi \vert T(r,\mathbold{\vartheta},\varphi)- T(1,\mathbold{\eta},\tau)\vert_2^d}\tilde{g}(\mathbold{\eta},\tau)  \mathrm{d}\tilde{S}(\mathbold{\eta}) \dd \tau,
\end{align*} for $r\in [0,1], \mathbold{\vartheta}\in [0,\pi]^{d-2}, \varphi\in [0,2\pi]$, where
\begin{align}
    \mathrm{d}\tilde{S}(\mathbold{\eta})=\sin(\eta_1)^{d-2}\sin(\eta_2)^{d-3}\cdots \sin(\eta_{d-2})\dd \eta_1\dots \dd \eta_d
\end{align}
 and $T$ denotes the coordinate transformation from spherical to Cartesian coordinates and $\tilde{g}$ the function $g$ expressed in the angular coordinates. For the sake of clarity, we have denoted the multi dimensional angular coordinates in bold Greek letters.
 
Since the partial differential equation we wish to study is given by the Laplace equation on the sphere, we can, similar to before, expand the solution by a series in term of the so-called  spherical harmonics. Spherical harmonics or $d$-spherical harmonics can be seen as the higher dimensional analog to the Fourier basis for $\rmL^2[0,1]$, see \cite{EftFry14SHPD}. They are defined as eigenfunctions of the eigenvalue problem to the Laplace--Beltrami operator on the $(d-1)$-sphere given by
\begin{align*}
  \Delta_{S^{d-1}} Y_{\ell,m}(\mathbold{\vartheta},\varphi)= l(2-d-l)Y^\ell_m(\mathbold{\vartheta},\varphi), \quad (\mathbold{\vartheta},\varphi)\in [0,\pi]^{d-2}\times [0,2\pi],
\end{align*} for $l\in \mathbb{N}$ and $m=1,\dots, N(d,l):=\frac{2l+d-2}{l} {l+d-3\choose l-1}$. The existence of the $d$-spherical harmonics has been proven in \cite{EftFry14SHPD}. Furthermore, it has also been shown that the set of all $d$-spherical harmonics form an orthonormal basis for $\rmL^2(S^{d-1})$ and that by employing spherical harmonic addition theorems, one can show  
\begin{align}\label{eq:addition.thm}
	\sum_{m=1}^{N(d,l)}|Y_{\ell,m}(\eta)|^2= \frac{N(d,l)}{\vert S^{d-1}\vert}=\frac{N(d,l)\Gamma(\frac{d}{2})}{2\pi^{\frac{d}{2}}} \quad \text{for all }(\mathbold{\vartheta},\varphi)\in [0,\pi]^{d-2}\times [0,2\pi],
\end{align}
 where $\vert S^{d-1}\vert$ denotes the surface area of the $d$-sphere. By the Cauchy--Schwarz inequality, for all $l,d \in \mathbb{N}$, we then obtain the inequality 
\begin{align} \label{estimate.addition}
     \sum_{m=1}^{N(d,l)}|Y_{\ell,m}(\eta)|&\leq N(d,l)^{\frac{1}{2}}\left(\sum_{m=1}^{N(d,l)}|Y_{\ell,m}(\eta)|^2\right)^{\frac{1}{2}} \notag\\
     &\leq N(d,l)^{\frac{1}{2}}\left(\frac{N(d,l)\Gamma(\frac{d}{2})}{2\pi^{\frac{d}{2}}}\right)^{\frac{1}{2}} \notag \\
     &=N(d,l)\left(\frac{\Gamma(\frac{d}{2})}{2\pi^{\frac{d}{2}}}\right)^{\frac{1}{2}}.
\end{align}  The $d$-spherical harmonics can be expressed in terms of Legendre polynomials \cite{EftFry14SHPD}, thus being polynomial-time computable for polynomial-time computable input. For examaple, the real-valued $3$-spherical harmonics are given by 
\begin{align}
Y_\ell^m( \theta , \varphi ) = \sqrt{\frac{(2\ell+1)}{4\pi} \frac{(\ell-m)!}{(\ell+m)!}} \, P_\ell^m ( \cos{\theta} ) \, e^{i m \varphi } 
\end{align} for $l=0,1,2,\dots$ and $m=-l,\dots,l$, where $P_\ell^m$ are the associated Legendre polynomials which in the closed form are given by
\begin{align*}
    P_\ell^m(x)=(-1)^{m} \cdot 2^{\ell} \cdot (1-x^2)^{m/2} \cdot  \sum_{k=m}^\ell \frac{k!}{(k-m)!}\cdot x^{k-m} \cdot \binom{\ell}{k} \binom{\frac{\ell+k-1}{2}}{\ell}.
\end{align*} However, apart from the polynomial-time computability, the exact form of the $d$-spherical harmonics is not required for our purposes. 

Having collected all tools, we are now in the position to state the main complexity result for the $d$-dimensional case. 
\begin{theorem} \label{th:Laplace.d}
	Let $r_0 \in \mathbb{R}_c\cap [0,1)$ be a polynomial-time computable number. We denote by  $S$ the solution operator that maps the input $g\in \rmC(S^{d-1})$ to the unique solution $u=Sg$ of the boundary value problem of the  Laplace equation \eqref{laplace}. Then, for all polynomial-time computable functions $g\in \rmC(S^{d-1})$, the mapping $[0,r_0]\times[0,\pi]^{d-2}\times[0,2\pi]\ni  (r,\mathbold{\eta},\varphi) \mapsto u(r,\mathbold{\eta},\varphi)=(Sg)(r,\mathbold{\eta},\varphi)$ is computable, and it is polynomial-time computable if and only if $FP_1=\# P_1$.
\end{theorem}
The proof is postponed to the Appendix \ref{se:appendix.b}.

\begin{rem} We wish to note that our computational complexity results for both Theorem \ref{th:Laplace.2d} and Theorem \ref{th:Laplace.d} are applicable to solutions on the open unit ball. Whether these results can be extended to the closed unit ball remains an open question.
\end{rem}

\section{The diffusion equation} \label{se:diffusion}
In this section, we focus on the one-dimensional diffusion equation. We distinguish between the case in which the spatial domain of the equations is given by a bounded set and the one in which it is given by an unbounded set. It is straight forward to see that for the bounded case, without loss of generality, we can choose the interval $[0,L], L>0,$ and for the unbounded case the interval $[0,+\infty)$. All other cases, except from $\mathbb{R}$, can be obtained by scaling and translation. In order to obtain a well-posed problem, we supplement the equation with appropriate Dirichlet or Neumann boundary conditions. In addition, we consider the equation with homogeneous and inhomogeneous right hand side. As we will see in the main results, the computational complexity of the solution will, depending on the boundary and initial conditions, be either in $\# P_1$ or in $\# P$. While the proof for $\# P_1$ is similar to the preceding proofs, the proof for $\# P$ is much more sophisticated and requires a deeper analysis of the solution. Furthermore, we want to remark that in most results regarding the $\# P$-complexity, we will only be able to show that $FP^\#P$ will be an upper bound for the computational complexity of the solution.  
\subsection{On a compact spatial domain}
Let $L>0$ and supplement the heat equation \eqref{heat} with initial and periodic boundary conditions given by 
\begin{align} \label{heat}
	\begin{cases}
		u_t(t,x)=\alpha  u_{xx}(t,x) \quad &\text{for } (t,x)\in (0,+\infty)\times (0,L),\\
		u(0,x)=g(x)\quad  &\text{for }  x\in  [0,L],\\
		u(t,0)=0=u(t,L)\quad  &\text{for } t\in (0,+\infty).
	\end{cases}
\end{align} Then, the unique solution of \eqref{heat} is given by
\begin{align}\label{sol.heat}
u(t,x)&=\frac{1}{\left(\sqrt{4\alpha \pi t} \right)^{1/2}}\int_0^L e^{-\frac{( y-x)^2}{4\alpha t}}g(y)\mathrm{d}y, \quad (t,x)\in [0,+\infty)\times [0,L],
\end{align} see \cite{Cann84ODHE}.
Now, we are in the position to state the main result:
\begin{theorem} \label{th:diffusion.compact}
Let $L>0$ and $t_0\in \mathbb{R}_c\cap (0,+\infty)$ be polynomial-time computable numbers. Denote by $S$ the solution operator that maps $g$ to the unique solution $u=Sg$ of the initial-boundary value problem of the heat equation \eqref{heat}. Then, for all polynomial-time computable functions $g\in \mathrm{C}([0,L])$, the mapping $[t_0,+\infty)\times [0,L]\ni (t,x)\mapsto u(t,x)=(Sg)(t,x)$ is computable, and it is polynomial-time computable if and only if $FP_1=\# P_1$.
\end{theorem} Since the proof has a similar structure as the proof of Theorem \ref{th:Laplace.d}, we postpone it to Appendix \ref{se:appendix.b}.

\subsection{On an unbounded spatial domain}
Next, we investigate the computational complexity of solutions to the inhomogeneous diffusion equation supplemented with inhomogeneous Dirichlet initial and boundary conditions in dimension one. More precisely, we show an upper bound for the complexity of solutions to the following Cauchy problem
\begin{align} \label{heat.0}
	\begin{cases}
		u_t(t,x)-\alpha u_{xx}(t,x)=f(x,t) \quad &\text{for } (t,x)\in (0,+\infty)\times [0,+\infty),\\
            u(0,x)=g(x)\quad  &\text{for }  x\in [0,+\infty),\\
		u(t,0)=h(t)\quad  &\text{for } t \in (0,+\infty),
	\end{cases}
\end{align} where $f:[0,+\infty)\times [0,+\infty)\rightarrow \mathbb{R}$. $g:[0,+\infty)\rightarrow \mathbb{R}$, and $h: [0,+\infty)\rightarrow \mathbb{R}$ are continuous functions. In fact, we will divide this Cauchy problem into three disjoint Cauchy problems by choosing  each time $f, g$, and $h$ non-zero and the other functions identical to zero. The reason for that is twofold. First, the proof is then divided into smaller parts and the solution to the general problem is then a superposition of the solution to each separate problem which easily follows from the linearity of the partial differential equation. Second, we will see that the computational complexity of each individual problem is not always the same. We will show that while for the solution corresponding to $g\neq 0$, we have $FP_1^{\# P_1}$ as an upper bound for the computational complexity, for solutions corresponding to $f\neq 0$ and $h\neq 0$, we obtain $FP^{\# P}$ as an upper bound for the computational complexity. 

\subsubsection{With non-homogeneous boundary condition $h\neq 0$}
First, we consider the diffusion equation with the following initial and boundary conditions 
\begin{align} \label{heat.2}
	\begin{cases}
		u_t(t,x)=\alpha u_{xx}(t,x) \quad &\text{for } (t,x)\in (0,+\infty)\times [0,+\infty),\\
            u(0,x)=0\quad  &\text{for }  x\in [0,+\infty),\\
		u(t,0)=h(t)\quad  &\text{for } t \in (0,+\infty),
	\end{cases}
\end{align} i.e., where $f=0$ and $g=0$. It can be shown that for a continuous function $h$, the solution is given by 
\begin{align}\label{sol.heat.2}
u(t,x)=\int_0^t\frac{x}{\sqrt{4\alpha \pi (t-s)^3}} e^{-\frac{x^2}{4\alpha (t-s)}}h(s)\mathrm{d}s, \quad (t,x)\in [0,+\infty)\times [0,+\infty), 
\end{align} see \cite{Cann84ODHE}. We note that the solution is obtained by convolving $h$ with the function 
\begin{align*}
    \Psi(t,x)=-2\alpha \partial_x \Phi(t,x)= \frac{x}{\sqrt{4\alpha \pi t^3}} e^{-\frac{x^2}{4\alpha t}},
\end{align*} where $\Phi(t,x)=\frac{1}{\sqrt{4\alpha \pi t}} e^{-\frac{x^2}{4\alpha t}}$ is the fundamental solution of the heat equation, i.e., it satisfies 
\begin{align*}
    \begin{cases}
		u_t(t,x)=\alpha u_{xx}(t,x) \quad &\text{for } (t,x)\in (0,+\infty)\times \mathbb{R},\\
            u(0,x)=\delta_0,
\end{cases}
\end{align*} with $\delta_0$ denoting the delta distribution concentrated in $x=0$.

The main complexity result reads as follows:

\begin{theorem} \label{th:diffusion.prl.h}
Let $x_0,x_1\in \mathbb{R}_c\cap (0,+\infty)$ be polynomial-time computable numbers. We denote by  $S$ the solution operator that maps $h$ to the unique solution $u=Sh$ of the initial-boundary value problem of the heat equation \eqref{heat.2}. Then, for all polynomial-time computable functions $h\in \mathrm{C}([0,+\infty))$ with $h(0)=0$, the mapping $[0,1]\times [x_0,x_1]\ni (t,x)\mapsto u(t,x)=(Sh)(t,x)$ is computable, and it is polynomial-time computable if $FP=\# P$.
\end{theorem} 

Similar as in the previous proofs, we want to find a sequence of approximations of the unique solution that can be computed by a Turing machine such that \eqref{computable.function} holds. Defining the function
\begin{align} \label{function.g}
g(t,x)=\frac{x}{t^{\frac{3}{2}}}e^{-\frac{x^2}{t}} , \quad (t,x)\in [0,+\infty)\times [0,+\infty),
\end{align} we note that 
\begin{align}\label{sol.equivalent}
u(t,x)=\frac{1}{\sqrt{\pi}}\int_0^t g(t-s,\frac{x}{\sqrt{4\alpha}})h(s)\mathrm{d}s, \quad (t,x)\in [0,+\infty)\times [0,+\infty).
\end{align} Together with Proposition \ref{taylor}, this shows that the function can be expanded in $t$ as a Taylor series and that the coefficients satisfy a growth condition. It allows us to rewrite the solution \eqref{sol.equivalent} within the convergence radius $\vert t-1\vert<1$ as
\begin{align}\label{sol.heat.2}
 u(t,x)=\frac{1}{\sqrt{\pi}}\sum_{n=0}^{\infty} \frac{g^{(n)}(1,\frac{x}{\sqrt{4\alpha}}) }{n!}\int_0^{t} (t-s-1)^n h(s)\mathrm{d}s, \quad x\in [0,+\infty).
\end{align} With this expression, we are ready to prove the main result.
\begin{proof}[Proof of Theorem \ref{th:diffusion.prl.h}] \textbf{Ad Computability:} Similar as in the previous theorems, we want to construct explicitly a sequence of functions whose computation is in $\#P$ and which approximates the exact solution \eqref{sol.heat.2} to the Cauchy problem \eqref{heat.2} in the sense of \eqref{computable.function}. Let $x\in \mathbb{R}_c\cap [0,+\infty)$ be fixed. We define 
\begin{align*}
    u_N(t,x):=\sum_{n=0}^{N^3} \frac{1}{\sqrt{\pi}}\frac{g^{(n)}(1,\frac{x}{\sqrt{4\alpha}}) }{n!}\int_0^{t-\frac{1}{N}} (t-s-1)^n h(s)\mathrm{d}s, 
\end{align*} for all $t\in [\frac{1}{N},1]$ and $u_N(t,x)$=0 for all $t\in [0,\frac{1}{N})$. Noting that  
\begin{align*}
    (t-s-1)^n=\sum_{k=0}^n {n \choose k} t^k (-s-1)^{n-k}=\sum_{k=0}^n p_{n-k}(s)t^k,
\end{align*} we obtain 
\begin{align*}
u_N(t,x)=\sum_{n=0}^{N^3} \frac{1}{\sqrt{\pi}}\frac{g^{(n)}(1,\frac{x}{\sqrt{4\alpha}}) }{n!} \sum_{k=0}^n p_{n-k}(s) \int_0^{t-\frac{1}{N}} t^k h(s)\mathrm{d}s.
\end{align*} $t\in [\frac{1}{N},1]$. We continue with showing that $u_N$ approximates $u$. It is easy to see that $u_N(0,x)=0=u(0,x)$ holds true. Now, let $t\in [\frac{1}{N},1]$. Then, we obtain 
\begin{align*}
   &\vert u(t,x)-u_N(t,x)\vert\\
   &=  \vert \int_0^t\frac{x}{\sqrt{4\alpha \pi (t-s)^3}} e^{-\frac{x^2}{4\alpha (t-s)}}h(s)\mathrm{d}s-  \sum_{n=0}^{N^3} \frac{1}{\sqrt{\pi}}\frac{g^{(n)}(1,\frac{x}{\sqrt{4\alpha}}) }{n!}\int_0^{t-\frac{1}{N}} (t-s-1)^n h(s)\mathrm{d}s\vert\\
   &=  \vert \int_{t-\frac{1}{N}}^{t}\frac{x}{\sqrt{4\alpha \pi (t-s)^3}} e^{-\frac{x^2}{4\alpha (t-s)}}h(s)\mathrm{d}s+ \int_0^{t-\frac{1}{N}}\frac{x}{\sqrt{4\alpha \pi (t-s)^3}} e^{-\frac{x^2}{4\alpha (t-s)}}h(s)\mathrm{d}s\\
   &\quad -  \sum_{n=0}^{N^3} \frac{1}{\sqrt{\pi}}\frac{g^{(n)}(1,\frac{x}{\sqrt{4\alpha}}) }{n!} \int_0^{t-\frac{1}{N}} (t-s-1)^n h(s)\mathrm{d}s\vert\\
  &\leq    \vert \int_{t-\frac{1}{N}}^{t}\frac{x}{\sqrt{4\alpha \pi (t-s)^3}} e^{-\frac{x^2}{4\alpha (t-s)}}h(s)\mathrm{d}s\vert\\
  &\quad+\vert \sum_{n=N^3+1}^{\infty} \frac{1}{\sqrt{\pi}}\frac{g^{(n)}(1,\frac{x}{\sqrt{4\alpha}}) }{n!} \int_0^{t-\frac{1}{N}} (t-s-1)^n h(s)\mathrm{d}s\vert\\
   &= I_1+I_2.
\end{align*} Now, we show that both terms $I_1$ and $I_2$ satisfy \eqref{computable.function}. Before, we continue estimating the term $I_1$, we note that the mapping $t\mapsto \frac{x}{\sqrt{4\alpha \pi t^3}} e^{-\frac{x^2}{4\alpha t}}$ is monotonically increasing on $[0,\frac{x}{\sqrt{6\alpha}}]$. Now, since $x\in [x_0,x_1]$, we choose $\tilde{N}_0=\tilde{N}_0(x_0,\alpha)\in \mathbb{N}$ independently of $t$ sufficiently large so that $\frac{1}{N}\leq \frac{x_0}{\sqrt{6\alpha}}$ for all $N\geq \tilde{N}_0$. Then, we obtain 
\begin{align*}
   I_1 &\leq   \int_{t-\frac{1}{N}}^{t} \frac{x}{\sqrt{4\alpha \pi (t-s)^3}} e^{-\frac{x^2}{4\alpha (t-s)}}\vert h(s)\vert \mathrm{d}s \\
   &=   \int_{0}^{\frac{1}{N}} \frac{x}{\sqrt{4\alpha \pi s^3}} e^{-\frac{x^2}{4\alpha s}}\vert h(t-s)\vert \mathrm{d}s \\
   &\leq \frac{x N^{\frac{3}{2}}}{\sqrt{4\alpha \pi}} e^{-\frac{x^2 N}{4\alpha}}  \int_{0}^{\frac{1}{N}}\vert h(t-s)\vert \mathrm{d}s\\
   &\leq \frac{x N^{\frac{1}{2}}}{\sqrt{4\alpha \pi}} e^{-\frac{x^2 N}{4\alpha}} \Vert h\Vert_{\rmC([0,2])}.
\end{align*} Now, we choose $\tilde{N}_1=\tilde{N}_1(x,\alpha,h)\in \mathbb{N}$, independently of $t$, sufficiently large so that $I_1\leq 2^{-N}$ for all $N\geq \tilde{N}_1$. For the term $I_2$, we recall that there exists some constant $C>0$ such that \eqref{derivative.bound} holds. We then obtain
\begin{align*}
   I_2 &\leq   \sum_{n=N^3+1}^{\infty} \frac{1}{\sqrt{\pi}}\left \vert \frac{g^{(n)}(1,\frac{x}{\sqrt{4\alpha}}) }{n!}\right \vert \int_0^{t-\frac{1}{N}} \vert t-s-1\vert ^n \vert h(s)\vert \mathrm{d}s\\
   &\leq   C\frac{x}{\sqrt{4\alpha \pi}}\sum_{n=N^3+1}^{\infty} (n+1) \int_0^{t-\frac{1}{N}} \vert t-s-1\vert ^n \vert h(s)\vert \mathrm{d}s\\
    &=   C\frac{x}{\sqrt{4\alpha \pi}}\sum_{n=N^3+1}^{\infty} (n+1) \int_{\frac{1}{N}}^t \vert s-1\vert ^n \vert h(t-s)\vert \mathrm{d}s\\
    &\leq  C \frac{x}{\sqrt{4\alpha \pi}}\sum_{n=N^3+1}^{\infty} (n+1) \left (1-\frac{1}{N}\right)^n \int_{\frac{1}{N}}^t  \vert h(t-s)\vert \mathrm{d}s\\
    &\leq  C \frac{x}{\sqrt{4\alpha \pi}}\Vert h\Vert_{\rmC([0,2])} \sum_{n=N^3+1}^{\infty} (n+1) \left (1-\frac{1}{N}\right )^n \\
    &\leq  C \frac{x}{\sqrt{4\alpha \pi}}\Vert h\Vert_{\rmC([0,2])}  \left (1-\frac{1}{N}\right )^{N^3+1} \left( N^4-N^5+N^2\right)  \\
    &\leq  C \frac{x}{\sqrt{4\alpha \pi}}\Vert h\Vert_{\rmC([0,2])}  \left (1-\frac{1}{N}\right )^{N^3+1}  (N^2+1)^2  \\
    &\leq C  \frac{x}{\sqrt{4\alpha \pi}}\Vert h\Vert_{\rmC([0,2])}  \left (1-\frac{1}{N}\right )^{N\cdot N^2} (N^2+1)^2\\
    &\leq  C \frac{x}{\sqrt{4\alpha \pi}}\Vert h\Vert_{\rmC([0,2])}  e^{- N^2} (N^2+1)^2,
\end{align*} where we used again the formula for the arithmetico-geometric series \eqref{arit.geom} and the fact that $\left (1-\frac{1}{N}\right )^N$ is strictly monotonically increasing in $N$ and converges to $e^{-1}$ from below as $N\rightarrow \infty$. Whereas the latter one is well-known, the former follows from the following reasoning: Showing that $y\mapsto \left (1-\frac{1}{y}\right)^y$ is monotonically increasing on $(1,+\infty)$ is equivalent to show that $y\mapsto \log \left (1-\frac{1}{y}\right)^y=y \log \left (1-\frac{1}{y}\right)$ is increasing. This in turn follows from the calculation:
\begin{align*}
    \frac{\rmd}{\rmd y}y \log \left (1-\frac{1}{y}\right)&=\log \left (1-\frac{1}{y}\right)+\frac{1}{y-1}\\
    &=\log(y-1)-\log(y)+\frac{1}{y-1}\\
    &=\int_{y-1}^y\left( \frac{1}{y-1}-\frac{1}{z}\right) \rmd z>0 \quad \text{for all }y>1.
\end{align*}
Hence, there exists another $\tilde{N}_2=\tilde{N}_2(x,\alpha,h)\in \mathbb{N}$, independently of $t$, such that $I_2\leq 2^{-N}$ for all $N\geq \tilde{N}_2$.
Finally, there exists some $N_3\in \mathbb{N}$, such that for all $t\in [0,\frac{1}{N}]$ and $N\geq N_3$
\begin{align*}
   \vert u(t,x)-u_N(t,x)\vert &=  \vert u(t,x)\vert \\
   &=\vert \int_0^t\frac{x}{\sqrt{4\alpha \pi (t-s)^3}} e^{-\frac{x^2}{4\alpha (t-s)}}h(s)\mathrm{d}s\vert \\
   &= \int_{0}^{t}\frac{x}{\sqrt{4\alpha \pi s^3}} e^{-\frac{x^2}{4\alpha s}}\vert h(t-s)\vert \mathrm{d}s \\
    &\leq \int_0^{\frac{1}{N}}\frac{x}{\sqrt{4\alpha \pi s^3}} e^{-\frac{x^2}{4\alpha s}}\vert h(t-s)\vert \mathrm{d}s \\
     &\leq \frac{x N^{\frac{3}{2}}}{\sqrt{4\alpha \pi}} e^{-\frac{x^2 N}{4\alpha}} \int_0^{\frac{1}{N}} \vert h(t-s)\vert\\
     &\leq\Vert h\Vert_{\rmC([0,2])}\frac{x N^{\frac{1}{2}}}{\sqrt{4\alpha \pi}} e^{-\frac{x^2 N}{4\alpha}},
\end{align*}  where we again made use of the fact that the mapping $t\mapsto \frac{x}{\sqrt{4\alpha \pi t^3}} e^{-\frac{x^2}{4\alpha t}}$ is strictly increasing on $[0,\frac{x}{\sqrt{6\alpha}}]$ and $N_3$ is to be chosen such that $\frac{1}{N_3}\leq \frac{x_0}{\sqrt{6\alpha}}$. Now, with the usual arguemnt, there exists $\tilde{N}_3=\tilde{N}_3(x,\alpha,h)\in \mathbb{N}$, independently of $t$, such that $\vert u(t,x)\vert\leq 2^{-N}$ for all $N\geq \tilde{N}_3$. Choosing $N_0=\max_{i=1,2,3}\tilde{N}_i$, we finally conclude that
\begin{align*} 
    \vert u(t,x)-u_N(t,x)\vert \leq  2^{-N} \quad \text{for all }(t,x)\in [0,1]\times [x_0,x_1] \text{ and }N\geq N_0,
\end{align*} which completes the proof.\\\\
\textbf{Ad Upper Bound:} We wish to note that according to Theorem \ref{Friedman1}, for all $k\in \mathbb{N}$ the functions $t\mapsto \int_0^t t^k h(s)\mathrm{d}s$ are polynomial-time computable if and only if $FP=\#P$ and the coefficients are all polynomial-time computable. Hence, for $x\in \mathbb{R}_c\cap[0,+\infty)$ the computation of the function $[0,1]\times [x_0,x_1]\mapsto u_N(t,x)$ remains in $FP^{\#P}$, and if $FP={\#P}$, then $FP^{\#P}=FP$, and the solution $u$ is polynomial-time computable.
 
\end{proof}

\begin{rem}
    We note that this complexity result is stronger than Theorem \ref{th:Laplace.2d}, \ref{th:Laplace.d}, and \ref{th:diffusion.compact} since $\# P_1$ is a subclass of $\#P$. That means that the calculation of  the map $t\mapsto u(t,x_0)=(Sg)(t,x_0)$ is computational more costly. This result also might indicate that the time dependence of the solution has in general higher complexity when there are time dependent boundary conditions or external forces.
\end{rem}

\subsubsection{With non-homogeneous external force $f\neq0 $}
First, we consider the one-dimensional diffusion equation with the initial and boundary conditions given by
\begin{align} \label{heat.3}
	\begin{cases}
		u_t(t,x)-\alpha u_{xx}(t,x)=f(t,x) \quad &\text{for } (t,x)\in (0,+\infty)\times [0,+\infty),\\
            u(0,x)=0\quad  &\text{for }  x\in [0,+\infty),\\
		u(t,0)=0\quad  &\text{for } t \in (0,+\infty).
	\end{cases}
\end{align} where for a continuous function $f:[0,\infty)\times [0,\times)\rightarrow \mathbb{R}$ the solution is given by 
\begin{align}\label{sol.heat.3}
u(t,x)=\int_0^t\int_0^\infty \frac{1}{\sqrt{4\alpha \pi (t-s)}} \left( e^{-\frac{(y-x)^2}{4\alpha (t-s)}}-e^{-\frac{(y+x)^2}{4\alpha (t-s)}}\right) f(y,s)\mathrm{d}y \mathrm{d}s 
\end{align} for all $(t,x)\in [0,+\infty)\times [0,+\infty)$, see \cite{Cann84ODHE}.

The following theorem states the main result.

\begin{theorem}\label{th:diffusion.prl.f}
Let $x_0,x_1, y_0\in \mathbb{R}_c\cap [0,+\infty)$ be polynomial-time computable numbers with $x_0>y_0$. We denote by $S$ the solution operator that maps $h$ to the unique solution $u=Sh$ of the initial-boundary value problem of the heat equation \eqref{heat.3}. Then, for all polynomial-time computable functions $f\in \mathrm{C}([0,+\infty)\times [0,+\infty))$ with $ f(\cdot,t)\in \mathrm{C}_c([0,y_0])$ and compact support in $[0,y_0]$ for all $t\in [0,1]$, the mapping $[0,1]\times [x_0,x_1]\ni (t,x)\mapsto u(t,x)=(Sh)(t,x)$ is computable, and it is polynomial-time computable if $FP=\# P$.
\end{theorem}

    The proofs of the previous two theorems reflect once again how much more complex the class $\# P$ is compared to the complexity class $\# P_1$. Unfortunately, the results do not show that the polynomial-time computability of solutions is sufficient for $FP=\# P$. However, we believe that this is the case and proving or disproving this claim is subject of future work. We formulate 
    \begin{con} Under the conditions of Theorems \ref{th:diffusion.prl.h} and \ref{th:diffusion.prl.f}, the solution is polynomial-time computable if and only if $FP=\# P$. 
    \end{con}
In the following section, we show that in fact the solution of the diffusion equation with space independent external force $f(t,x)=f(t)$ is indeed $\# P$-complete. However, as we will see this assumptions leads to a space independent solution that is a solution to a trivial ordinary differential equations.

\subsubsection{With non-homogeneous boundary condition $g\neq 0$}
Next, we consider the diffusion equation with the initial and boundary conditions given by 
\begin{align} \label{heat.4}
	\begin{cases}
		u_t(t,x)=\alpha u_{xx}(t,x) \quad &\text{for } (t,x)\in (0,+\infty)\times [0,+\infty),\\
            u(0,x)=g(x)\quad  &\text{for }  x\in [0,+\infty),\\
		u(t,0)=0\quad  &\text{for } t \in (0,+\infty),
	\end{cases}
\end{align} i.e., where $f=0$ and $h=0$. For continuous $g:[0,+\infty)\rightarrow \mathbb{R}$, the solution is given by 
\begin{align}\label{sol.heat.4}
u(t,x)=  \frac{1}{\sqrt{4\alpha \pi t}} \int_0^\infty  \left( e^{-\frac{(y-x)^2}{4\alpha (t-s)}}-e^{-\frac{(y+x)^2}{4\alpha (t-s)}}\right) h(y)\mathrm{d}y \mathrm{d}s 
\end{align} for all $(t,x)\in [0,+\infty)\times [0,+\infty)$, see \cite{Cann84ODHE}.
As we will see in the next statement of the main result, unlike in the previous results, the upper bound for the computational complexity of the solution is $FP^{\# P_1}$. 
\begin{theorem}\label{th:diffusion.prl.g}
Let $t_0\in (0,+\infty)$ be a polynomial-time computable number and denote with $S$ the solution operator that maps $f$ to the unique solution $u=Sf$ of the initial-boundary value problem of the heat equation \eqref{heat.4}. Then, for all polynomial-time computable functions $g\in \mathrm{C}([0,+\infty))$, the mapping $[0,t_0]\times [0,+\infty) \ni t \mapsto u(t,x)=(Sf)(t,x)$ is computable for all $x\in [0,+\infty)$, and it is polynomial-time computable if $FP_1=\# P_1$.
\end{theorem} 

\begin{proof}
     First, we choose $g\in \rmC_c([a,b])$, i.e., being compactly supported in $[a,b]$ with $[a,b]\subsetneq [0,1]$, such that the computation of $\int_0^1 g(y)\dd y$ is in $\# P_1$ according to Theorem \ref{Friedman1}, and $g$ is zero outside of $[a,b]$ which can be achieved by interpolation. Since the expression of the solution has a similar structure as in the case for non-homogeneous external force, i.e., Theorem \eqref{th:diffusion.prl.f}, the proof follows along the same lines.
\end{proof}

Putting all the previous results together, we have shown that the computational complexity of solutions to the general Cauchy problem \eqref{heat.0} lie in $\# P_1$.

\subsubsection{With external force $f\neq0 $ and homogeneous Neumann boundary condition}
Finally, we consider the diffusion equation with non-vanishing external force $f$ and homogeneous initial and Neumann boundary conditions given by 
\begin{align} \label{heat.5}
	\begin{cases}
		u_t(t,x)-\alpha u_{xx}(t,x)=f(t,x) \quad &\text{for } (t,x)\in (0,+\infty)\times [0,+\infty),\\
            u(0,x)=0\quad  &\text{for }  x\in [0,+\infty),\\
		u_x(t,0)=0\quad  &\text{for } t \in (0,+\infty),
	\end{cases}
\end{align} where for a continuous function $f:[0,+\infty)\times [0,+\infty)\rightarrow \mathbb{R}$ the solution is given by 
\begin{align*}
u(t,x)&=\int_0^t\int_0^\infty \frac{1}{\sqrt{4\alpha \pi (t-s)}} \left( e^{-\frac{(y-x)^2}{4\alpha (t-s)}}+e^{-\frac{(y+x)^2}{4\alpha (t-s)}}\right) f(s,y)\mathrm{d}y \mathrm{d}s
\end{align*} for all $(t,x)\in [0,+\infty)\times [0,+\infty)$. Assuming that $f(x,t)=f(t)$ for all $(t,x)\in [0,+\infty)\times [0,+\infty)$, we obtain 
\begin{align*}
u(t,x)=\int_0^t\int_{-\infty}^\infty \frac{1}{\sqrt{4\alpha \pi (t-s)}} e^{-\frac{(y-x)^2}{4\alpha (t-s)}} f(s)\mathrm{d}y \mathrm{d}s =\int_0^t  f(s) \mathrm{d}s
\end{align*} for all $(t,x)\in [0,+\infty)\times [0,+\infty)$, where we used the formula
\begin{align*}
    \int_{\infty}^\infty \frac{1}{\sqrt{4\alpha \pi (t-s)}} e^{-\frac{(y-x)^2}{4\alpha (t-s)}} \mathrm{d}y=1
\end{align*} for all $x,s,t\in \mathbb{R}$ with $t>s$, see \cite{Cann84ODHE}. Then, Theorem \ref{Friedman1} ensures the existence of an polynomial-time computable function $f$ such that the computation of $t\mapsto u(t)$ is $\# P$-complete. Thus, we have 

\begin{theorem}\label{th:diffusion.neumann}
Let $t_0\in (0,+\infty)$ be polynomial-time computable number and denote with $S$ the solution operator that maps $f$ to the unique solution $u=Sf$ of the initial-boundary value problem of the heat equation \eqref{heat.5}. Then, for all polynomial-time computable functions $f\in \mathrm{C}([0,+\infty))$, the mapping $[0,t_0] \ni t \mapsto u(t,x)=(Sf)(t,x)$ is computable for all $x\in [0,+\infty)$, it is polynomial-time computable if and only if $FP=\# P$.
\end{theorem} Inserting the solution into the diffusion equation, we find that the partial differential equation 
\begin{align} \label{heat.6}
\begin{cases}
	u_t(t,x)=f(t) \quad &\text{for } (t,x)\in (0,+\infty)\times [0,+\infty),\\
            u(0,x)=0\quad  &\text{for }  x\in [0,+\infty),
	\end{cases}
\end{align} has become a trivial ordinary differential equation with no diffusion term. Nevertheless, our results clearly show that the solution to the inhomogeneous diffusion equation possesses solutions that are $\#P$-complete. However, whether this is the case for an  external force that is not constant in space, is not clear as it is well known that the Laplace operator has a smoothing effect on the solution meaning that the solution is infinitely many times differentiable in space-time, and under stronger conditions even analytic for $t>0$. Therefore, it is possible that the space dependence of the solution makes it difficult to show the $\# P$-completeness of solutions to \eqref{heat.3}. However, it is still not clear what effect the Laplace operator precisely has on the complexity of the solutions. So far it has been only established that if a computable function $f:[a,b]\rightarrow \mathbb{R}$ belongs to $C^2([a,b])$, then its first derivative is computable, see \cite{PouRic89CIAP}. In particular, if $f\in C^\infty([a,b])$, then all its derivatives are computable. On the other hand, it has also been shown that there exists twice differentiable functions whose first derivative is continuous but not computable. This is due to the fact that computability necessitates effective uniform continuity, see \cite{Myhi71RFDN,PouRic89CIAP}.

\section{Conclusion and Outlook} \label{se:conclusion}
In this article, we proved that the computational complexity of solutions to a class of Cauchy problems for the Laplace equation and the diffusion equation with polynomial-time computable input data are in $\# P_1$ and $\# P$, respectively. More specifically, we showed that the time dependence of the solutions for the corresponding input data with polynomial complexity, are $\#P$-complete whereas the space dependence is $\# P_1$-complete depending on the initial and boundary conditions as well as the shape of the domain. This implies that if $FP_1\neq \# P_1$ and $FP\neq \# P$, then the complexity of the solution leads to a complexity blowup meaning that the computation time for obtaining an approximation up to an error of $2^{-n}$, which corresponds to $n$ significant digits on a digital computer, grows faster than any polynomial in $n$. However, it's important to note that some of our computational complexity results, especially those related to $\#P_1$, are valid only for solutions close to the boundary. For instance, in Theorem \ref{th:Laplace.2d} and \ref{th:Laplace.d}, we establish the optimality result for solutions of the Laplace equation on any ball with a radius $r_0<1$. Additionally, in Theorem \ref{th:diffusion.prl.g}, we obtain the hardness result for solutions of the heat equation after an initial time $t_0>0$. Since, in both cases, the solutions evaluated at $r=1$ and $t=0$, respectively, are polynomial-time computable due to the chosen boundary and initial conditions, we hypothesize that our results can be extended to the boundary. Nevertheless, this remains an open question.  

We would also like to note that our computational complexity result for the solution operator is understood in a non-uniform sense. This means that we fix an initial or boundary condition, in contrast to \cite{KawCoo10CTOA}, where operator complexity is introduced. Nevertheless, our results indicates, that the solution operator itself has inherently high complexity and that there exists no numerical scheme that can compute the solutions efficiently. As a consequence, it would be computationally very costly to obtain a reasonably good approximation of the solution despite the efficiency of the numerical scheme as the complexity is inherent to the solution operator. Hence, even simple linear partial differential equations with constant coefficients have solutions whose calculations on a digital computer are very costly, if it can be calculated at all. In fact, many methods that are used to solve partial differential equations, e.g., numerical methods like the Galerkin scheme or the Euler scheme, transformations like the Fourier transform or the  Laplace transform, or methods for solving optimization problems contain operations like integration or differentiation that lead in the calculation of solutions on a digital computer to either a complexity-blowup or a non-computability, see \cite{Ko82NRCC,Frie84CCMI, DuKo89CCID,KoLin95CMOP,BocPohACFS, BocPoh20ACHT, BocPoh20ASSF, BocPoh20TMCT, BoFoKu22LDLI, BoFoKu22IPSP}. It is also interesting to note that the Boson sampling problem in quantum computing, more specifically linear optics, has the same computational complexity as computing an integral \cite{AarArk13CCLO}.

Furthermore, since the physical phenomena described by the Laplace and the diffusion equation is encoded in the solution operator, we showed that physical phenomena, in general, have intrinsically high complexity that can exactly be captured by the Turing machine. Thus, predicting the values of solutions to the instationary and steady-state heat equation for polynomial-time input data to finite accuracy, cannot be done by a classical computer in polynomial-time, unless $FP= \#P$. 

It would be interesting to study the computational complexity of solutions to the Laplace and the heat equation based on other computing models, e.g. analog computing models where operations that cause the high complexity or non-computability like the Fourier transform, integration, or differentiation would be then obtained by measuring physical quantities, see, e.g., \cite{SMCG14PMOM} where the authors realise the mathematical operations like the differentiation or the Fourier transform with metamaterials. However, one has to take into account errors in the process of measuring as well as fundamental physical limitations like the Heisenberg's uncertainty principle which does not allow infinite precision in the simultaneous measuring of different observables. Studying solutions on analog or quantum computing models like the Blum–Shub–Smale machine is especially intriguing given the significant scientific interest in quantum computers, biocomputing, and neuromorphic computing, as evidenced by \cite{BlShSm89CCRN,BoFoKu22LDLI, BoFoKu22IPSP}.

Despite the open questions and problems we have already formulated, it is also of great importance to quantify the data that leads to complexity-blowups or non-computability of solutions. Unfortunately, obtaining a satisfying answer to that question is at least as hard as solving the $P$ vs. $NP$ problem, since it would quantify the sets $FP$ and $FP_1$ in relation to $\# P$ and $\# P_1$, respectively.

Since the solution of the large majority of partial differential equations, especially nonlinear partial differential equations, do not have a closed form, a further potential direction is to investigate the computational complexity of solution to certain nonlinear partial differential equations, e.g., the Navier--Stokes equtions in fluid dynamics, the Black--Scholes equation in finance, or the Hamilton--Jacobi--Bellman equation from optimal control theory. This requires good a priori estimates of the solution as it has been done for ordinary differential equations, see Section \ref{se:ODE}. A more feasible aspect to investigate is the computational complexity of common numerical schemes that compute an approximation of a solution. However, this would only give the computational complexity of the numerical scheme and not the computational complexity of the solutions as we showed in this paper.

\appendix
\section{Appendix} \label{se:appendix}
In this section, we provide some analytical tools and results that are used in the article. First, we give an explicit expression of an arithmetico-geometric series.
\begin{pro}  \label{le:polynome}
Let $m,p\in \mathbb{N}$ and $x\in (-1,1)$. Then, there exists a polynomial $P_{2p}$ of degree $2p$ such that 
\begin{align} \label{polynome}
\sum_{k=m}^\infty x^k \frac{(k+p)!}{k!}=\frac{x^{m}}{(1-x)^{p+1}}P_{2p}(x,m)
\end{align}
\end{pro}
\begin{proof} It is easy to see that the following power series is absoluteley convergent
	\begin{align*}
\sum_{k=m}^\infty x^k=\frac{x^m}{1-x}.
	\end{align*} Thus by analyticity, all derivatives are analytic and the derivatives are given by 
\begin{align*}
\frac{\dd^{p}}{\dd x^{p}}	\frac{x^{p+m}}{1-x}=\frac{\dd^{p}}{\dd x^{p}}\sum_{k=m}^\infty x^{k+p}=\sum_{k=m}^\infty \frac{\dd^{p}}{\dd x^{p}}x^{k+p}=\sum_{k=m}^\infty x^k \frac{(k+p)!}{k!}.
\end{align*}  Now, we prove the statement by induction over the number of derivatives $p$. For $p=1$, we obtain 
\begin{align*}
    \frac{\dd}{\dd x} \frac{x^{1+m}}{1-x}=  \frac{(1+m)x^{m}(1-x)+x^{1+m}}{(1-x)^2}=\frac{x^{m}}{(1-x)^2}((1+m)(1-x)+x).
\end{align*} By induction, let \eqref{polynome} hold true for $p\in \mathbb{N}$. We calculate 
\begin{align*}
   \frac{\dd^{p+1}}{\dd x^{p+1}}	\frac{x^{p+1+m}}{1-x}&= \frac{\dd}{\dd x} \frac{x^{m+1}}{(1-x)^{p+1}}P_{2p}(x,m+1)\\
   &=\frac{((m+1) x^{m}P_{2p}(x,m+1)+x^{m+1} \partial_x P_{2p}(x,m+1))(1-x)^{p+1}}{(1-x)^{2p+2}}\\
   &\quad+ \frac{(x^{m+1} P_{2p}(x,m+1)(p+1)(1-x)^{p}}{(1-x)^{2p+2}}{(1-x)^{2p+2}}\\
   &=\frac{x^{m}}{(1-x)^{p+2}}(((m+1) P_{2p}(x,m+1)+x \partial_x P_p(x,m+1))(1-x))\\
   &\quad + \frac{x^{m}}{(1-x)^{p+2}}(x P_{2p}(x,m+1)(p+1))\\
   &= \frac{x^{m}}{(1-x)^{p+2}}P_{2(p+1)}(x,m)
\end{align*} for a polynomial $P_{2(p+1)}$ of degree $2(p+1)$ and thus the completion of the proof.
\end{proof}
If we take a closer look at the proof, we have proven the following formula:
\begin{cor}  \label{cor:polynome}
Let $m\in \mathbb{N}$ and $x\in (-1,1)$. Then, we have 
\begin{align}\label{arit.geom}
    \sum_{k=m}^{\infty} (k + 1) x^k = \frac{(x^m (m (1-x) + 1))}{(x - 1)^2} 
\end{align} for all $\vert x\vert <1$ and $m \in \mathbb{N}$.
\end{cor}

The following result shows that the function $g:[0,2]\times \mathbb{R}$ given by
\begin{align} \label{function.g}
g(t,x)=\frac{x}{t^{\frac{3}{2}}}e^{-\frac{x^2}{t}} , \quad (t,x)\in [0,+\infty)\times [0,+\infty),
\end{align} can be expanded in $t$ by a Taylor series and that its time derivative grows linearly in the number of derivatives as well as in space. 
\begin{pro} \label{taylor} Let $g:[0,2]\times [0,+\infty)\rightarrow \mathbb{R}$ be defined as in \eqref{taylor}. Then, the function has a Taylor expansion around $t_0=1$ with convergence radius $R=1$ given by
\begin{align}
g(t,x)=\sum_{n=0}^\infty \frac{g^{(n)}(1,x)}{n!}(t-1)^n \quad (t,x)\in (0,2)\times [0,+\infty),
\end{align} where 
\begin{align}\label{derivative}
g^{(n)}(t,x)=xe^{-\frac{x^2}{t}}\sum_{m=0}^n \frac{x^{2(n-m)}}{t^{\frac{3}{2} +2n-m}} (-1)^m {n\choose m} \frac{\Gamma(\frac{3}{2}+n)}{\Gamma(\frac{3}{2}+n-m)}.
\end{align}
Furthermore, for all polyonmial-time computable $x_0\in  \mathbb{R}_c\cap [0,+\infty)$, the function $ \mathbb{R}_c\cap [0,x_0] \ni x\mapsto g^{(n)}(1,x)$ is polynomial-time computable and there exists some constant $C=C(x_0)>0$ such that 
\begin{align}
\label{derivative.bound}
\frac{\vert g^{(n)}(1,x)\vert}{n!} \leq C(n+1)x \quad \text{for all }x\in [0,x_0], n\in \mathbb{N}.
\end{align}
\end{pro}
\begin{proof}
Since for every $x\in [0,+\infty)$ the function $g$ is analytic on $\mathbb{R}\lbrace 0\rbrace$, it follows from complex analysis. We first prove the Formula \eqref{derivative} by induction over the number of derivatives. It is easy to check that $g$ has a Taylor expansion in $t$ and the radius of convergence of the power series f centered on a point $t=1$ is equal to the distance from $t=1$ to the nearest singularity where $g$ cannot be analytically continued which is $t=0$. Hence the radius of convergence is $R=1$. \\ Next, we prove formula \eqref{derivative} for the derivative of $g$ with respect to the time variable by induction over the number of derivatives. We have

\begin{align*}
&g^{(0)}(t,x)=\frac{x}{t^{\frac{3}{2}}}e^{-\frac{x^2}{t}}=g(t,x),\\
&g^{(1)}(t,x)=\left(\frac{x^2}{t^{\frac{3}{2}+2}}-\frac{\frac{3}{2}}{t^{\frac{3}{2}+1}}\right)xe^{-\frac{x^2}{t}}=\partial_t g(t,x).\\
\end{align*} We assume that \eqref{derivative} is correct for $k=0,\dots,n$. This yields 
\begin{align*}
\partial_t g^{(n)}(t,x) &=xe^{-\frac{x^2}{t}}\sum_{m=0}^n \frac{x^{2(n+1-m)}}{t^{\frac{3}{2} +2(n+1)-m}} (-1)^m {n\choose m} \frac{\Gamma(\frac{3}{2}+n)}{\Gamma(\frac{3}{2}+n-m)}\\
&\quad + xe^{-\frac{x^2}{t}}\sum_{m=0}^n \left(\frac{3}{2} +2n-m\right)\frac{x^{2(n-m)}}{t^{\frac{3}{2} +2n-m+1}} (-1)^{m+1} {n\choose m} \frac{\Gamma(\frac{3}{2}+n)}{\Gamma(\frac{3}{2}+n-m)}\\
&=xe^{-\frac{x^2}{t}}\sum_{m=1}^n \frac{x^{2(n+1-m)}}{t^{\frac{3}{2} +2(n+1)-m}} (-1)^m {n\choose m} \frac{\Gamma(\frac{3}{2}+n)}{\Gamma(\frac{3}{2}+n-m)}\\
&\quad + xe^{-\frac{x^2}{t}}\sum_{m=0}^{n-1} \left(\frac{3}{2} +2n-m\right)\frac{x^{2(n-m)}}{t^{\frac{3}{2} +2n-m+1}} (-1)^{m+1} {n\choose m} \frac{\Gamma(\frac{3}{2}+n)}{\Gamma(\frac{3}{2}+n-m)}\\
&\quad + xe^{-\frac{x^2}{t}}\left(\frac{x^{2(n+1)}}{t^{\frac{3}{2} +2(n+1)}}+ \frac{\left(\frac{3}{2} +n\right)}{t^{\frac{3}{2} +n+1}} (-1)^{n+1}  \frac{\Gamma(\frac{3}{2}+n)}{\Gamma(\frac{1}{2})} \right)\\
&=xe^{-\frac{x^2}{t}}\sum_{m=0}^{n-1} \frac{x^{2(n-m)}}{t^{\frac{3}{2} +2n -m+1}} (-1)^{m+1} {n\choose m+1} \frac{\Gamma(\frac{3}{2}+n)}{\Gamma(\frac{3}{2}+n-m-1)}\\
&\quad + xe^{-\frac{x^2}{t}}\sum_{m=0}^{n-1} \left(\frac{3}{2} +2n-m\right)\frac{x^{2(n-m)}}{t^{\frac{3}{2} +2n-m+1}} (-1)^{m+1} {n\choose m} \frac{\Gamma(\frac{3}{2}+n)}{\Gamma(\frac{3}{2}+n-m)}\\
&\quad + xe^{-\frac{x^2}{t}}\left(\frac{x^{2(n+1)}}{t^{\frac{3}{2} +2(n+1)}}+ (-1)^{n+1}  \frac{\Gamma(\frac{3}{2}+(n+1))}{\Gamma(\frac{1}{2}) t^{\frac{3}{2} +n+1}} \right)\\
&=xe^{-\frac{x^2}{t}}\sum_{m=0}^{n-1} \frac{x^{2(n-m)}}{t^{\frac{3}{2} +2n -m+1}} (-1)^{m+1} a^n_m \frac{\Gamma(\frac{3}{2}+n)}{\Gamma(\frac{3}{2}+n-m)}\\
&\quad + xe^{-\frac{x^2}{t}}\left(\frac{x^{2(n+1)}}{t^{\frac{3}{2} +2(n+1)}}+ (-1)^{n+1}  \frac{\Gamma(\frac{3}{2}+(n+1))}{\Gamma(\frac{1}{2}) t^{\frac{3}{2} +n+1}} \right), 
\end{align*} 
where 
\begin{align*}
    a^n_m &= \left( {n\choose m+1}\left(\frac{3}{2}+n-m-1\right)+{n\choose m}\left(\frac{3}{2} +2n-m\right)  \right)\\
    &=\left( {n\choose m+1}+{n\choose m}\right) \left(\frac{1}{2}+n-m\right) +{n\choose m}\left(n+1\right)  \\
    &= {n+1\choose m+1}\left(\frac{1}{2}+n-m\right) +{n+1\choose m+1}\left(m+1\right)\\
     &= {n+1\choose m+1}\left(\frac{1}{2}+n+1\right).
\end{align*} Since $\Gamma(\frac{3}{2}+n+1)=\Gamma(\frac{3}{2}+n)(\frac{1}{2}+n+1)$, this shows $\partial_t g^{(n)}(t,x)=g^{(n+1)}(t,x)$ and thus the desired formula \eqref{derivative} for all $n\in \mathbb{N}$. 

In order to show the uniform bound of the derivates \eqref{derivative.bound}, we note that 

\begin{align*}
   g^{(n)}(1,x) &= xe^{-x^2}\sum_{m=0}^n x^{2(n-m)} (-1)^m {n\choose m} \frac{\Gamma(\frac{3}{2}+n)}{\Gamma(\frac{3}{2}+n-m)}  \\
    &= xe^{-x^2} \Gamma\left(\frac{3}{2}+n\right)(-1)^n \sum_{m=0}^n x^{2m} (-1)^m {n\choose m}\frac{1}{\Gamma(\frac{3}{2}+m)}\\
    &= \frac{xe^{-x^2} \Gamma\left(\frac{3}{2}+n\right) 2 (-1)^n}{\sqrt{\pi}}  M\left(-n,\frac{3}{2},x^2\right) ,
\end{align*} where $M$ is the so-called confluent hypergeometric function, i.e., the mapping $z\mapsto w(z)=M(-n, \frac{3}{2}, z)$ satisfies the differential equation
\begin{align*}
    z w''(z)+\left(\frac{3}{2}-z\right)w'(z)+nw(z)=0,
\end{align*} see \textsc{Abramowitz} and \textsc{Stegun} \cite[Chapter 13, pp. 504]{AbrSte64HMFF}. 

By \cite[Formula 13.5.14]{AbrSte64HMFF}, the hypergeometric function has the following asymptotic behaviour 
\begin{align*}
M(-n,a,x^2)&=\Gamma\left(a\right) \pi^{-\frac{1}{2}}e^{\frac{1}{2}x^2}((2a+4n)x^2)^{\frac{1}{4}-\frac{1}{2}a}\cos\left((2ax^2+4nx^2)^{\frac{1}{2}}-\left(\frac{1}{2}a-\frac{1}{4}\right)\pi\right)\\
&\quad\cdot\left [1+\mathcal{O}\left (\left \vert \frac{1}{2}a+n\right \vert^{-\frac{1}{2}}\right)\right]
\end{align*}  as $n\rightarrow \infty$ for $a\in \mathbb{C}$ bounded and $x\in \mathbb{R}$. Hence, for all $x_0\in [0,+\infty)$ there exists some constant $C=C(x_0)>0$ independently of $n$ such that 
\begin{align*} 
  \frac{\vert g^{(n)}(1,x)\vert}{n!} &= \frac{xe^{-x^2} \Gamma\left(\frac{3}{2}+n\right) 2}{n! \sqrt{\pi}} \left \vert M\left(-n,\frac{3}{2},x^2\right)\right \vert \\
  &\leq C \frac{xe^{-x^2} \Gamma\left(\frac{3}{2}+n\right) }{n!} \left(1+e^{\frac{1}{2}x^2} \left(\frac{3}{4}+n\right)^{-\frac{1}{2}} \frac{\vert \cos\left((3+4n)^{\frac{1}{2}}x-\frac{1}{2} \pi\right)\vert }{x}\right)\\
   &= C \frac{xe^{-x^2} \Gamma\left(\frac{3}{2}+n\right) }{n!} \left(1+e^{\frac{1}{2}x^2} \left(\frac{3}{4}+n\right)^{-\frac{1}{2}} \vert \sin\left((3+4n)^{\frac{1}{2}}\xi-\frac{1}{2} \pi\right)(3+4n)^{\frac{1}{2}} \vert\right)\\
    & \leq C \frac{xe^{-x^2} \Gamma\left(\frac{3}{2}+n\right) }{n!} \left(1+\frac{1}{2}e^{\frac{1}{2}x^2} \right)\\
    & \leq C \frac{x \Gamma\left(\frac{3}{2}+n\right) }{n!} \left(e^{-x^2}+1 \right)\\
     & \leq C \frac{2x (n+1)! }{n!}\\
     &= C 2x (n+1),
\end{align*} where the existence of $\xi\in (0,x)$ follows from the mean value theorem and we used the fact that $\Gamma(n+2)=(n+1)!$. This proves the bound \eqref{derivative.bound}.

Finally, let $x\in \mathbb{R}_c\cap [0,+\infty)$. Then, it is easy to see that the value 
\begin{align*}
    g^{(n)}(1,x)=e^{-x^2}\sum_{m=0}^n x^{2(n-m)} (-1)^m {n\choose m} \frac{\Gamma(\frac{3}{2}+n)}{\Gamma(\frac{3}{2}+n-m)}, 
\end{align*} is as a composition of a finite product and a finite sum of polynomial-time computable numbers also polynomial-time computable.
\end{proof}
As a corollary, we find the following result.

\begin{cor}\label{taylor.cor}
Let $\tilde{g}:[0,+\infty)\times [0,+\infty)\rightarrow \mathbb{R}$ with 
\begin{align*} 
\tilde{g}(t,x)=\frac{1}{t^{\frac{1}{2}}}e^{-\frac{x^2}{t}} , \quad (t,x)\in [0,+\infty)\times [0,+\infty),
\end{align*} be given. Then, the function $\tilde{g}$ has a Taylor expansion around $t_0=1$ with convergence radius $R=1$ given by
\begin{align}\label{taylor.g2}
\tilde{g}(t,x)=\sum_{n=0}^\infty \frac{\tilde{g}^{(n)}(1,x)}{n!}(t-1)^n \quad (t,x)\in (0,2)\times [0,+\infty),
\end{align} where 
\begin{align}\label{derivative.2}
\tilde{g}^{(n)}(t,x)=\frac{g^{(n)}(t,x) t+g^{(n-1)}(t,x)}{x},
\end{align} where $g^{(n)}$ is given by \eqref{derivative}. Furthermore, for all polynomial-time computable $x_0\in  \mathbb{R}_c\cap [0,+\infty)$, the function $\mathbb{R}_c\cap [0,x_0] \ni x\mapsto \tilde{g}^{(n)}(1,x)$ is polynomial-time computable and there exists some constant $C=C(x_0)>0$ such that 
\begin{align}
\label{derivative.bound.g2}
\frac{\vert \tilde{g}^{(n)}(1,x)\vert}{n!} \leq C(n+1)\quad \text{for all }x\in [0,x_0], n\in \mathbb{N}.
\end{align}
\end{cor}
\begin{proof} This follows immediately from the fact that $\tilde{g}(t,x)=\frac{t}{x}g(t,x)$ for all $(t,x)\in [0,+\infty)\times [0,+\infty)$ and Proposition \ref{taylor}.
\end{proof}

\section{Appendix} \label{se:appendix.b}
In this section, we present the proofs of Theorem \ref{th:Laplace.d} and \ref{th:diffusion.compact} as well as Theorem \ref{th:diffusion.prl.f} as their structure is similar to the proofs of Theorem \ref{th:Laplace.2d} and Theorem \ref{th:diffusion.prl.h} but require different tools. First, we present the proof of Theorem \ref{th:Laplace.d}.

\begin{proof}[Proof of Theorem \ref{th:Laplace.d}] \textbf{Ad Computability:} Similar as before, the boundary value problem to the Laplace equation \eqref{laplace} is uniquely solvable and the solution can be expressed in terms of the spherical harmonics by 
\begin{align}\label{sol.harmonics2}
	u(r, \mathbold{\eta}, \varphi) = \sum_{\ell=0}^\infty \sum_{m=1}^{N(d,l)} c_{\ell,m} r^\ell Y_{\ell,m} (\mathbold{\eta}, \varphi), \quad (r,\mathbold{\eta},\varphi) \in [0,1]\times [0,\pi]^{d-2}\times [0,2\pi],
\end{align} where
\begin{align*}
	c_{\ell,m}= \int_{S^{d-1}} Y_{\ell,m} (\mathbold{\eta}) g (\mathbold{\eta} ) \dd S^{d-1}, \quad  l\in \mathbb{N}, l=1, \dots, N(d,l),
\end{align*} see \cite{EftFry14SHPD}. Evaluating the solution $u$ on the boundary, we obtain by definition of a solution
\begin{align}
u(1, \mathbold{\eta}, \varphi)=g(\mathbold{\vartheta}), \quad \mathbold{\vartheta}\in S^{d-1},
\end{align}
where $\mathbold{\vartheta}=(\mathbold{\eta},\varphi)$. It is easy to check that the series converges for all $r\leq 1$ and that by Parseval's Identity, we obtain 
 \begin{align*}
\int_{S^{d-1}}  \vert g (\mathbold{\vartheta})\vert^2 \dd S^{d-1} = \sum_{\ell=0}^\infty \sum_{m=1}^{N(d,l)}\vert c_{\ell,m}\vert^2 <+\infty,
\end{align*} which in turn implies 
\begin{align}\label{est.3}
\sup_{l,m\in \mathbb{N}_0} \vert c_{l}^m\vert \leq  \left( \sum_{\ell=0}^\infty \sum_{m=1}^{N(d,l)}\vert c_{\ell,m}\vert^2 \right)^{\frac{1}{2}} =\left(\int_{S^{d-1}} \vert g (\mathbold{\vartheta})\vert^2 \dd S^{d-1}\right)^{\frac{1}{2}} \leq \Vert g\Vert_{\infty} \left(\frac{2\pi^{\frac{d}{2}}}{\Gamma(\frac{d}{2})}\right)^{\frac{1}{2}}.
\end{align} In the exact same manner as in \eqref{eq:partial.sum.laplace}, we define 
\begin{align} 	
u_N(r, \mathbold{\eta}, \varphi) := \sum_{\ell=0}^{N K} \sum_{m=1}^{N(d,l)} c_{\ell,m} r^\ell Y_{\ell,m} (\mathbold{\eta}, \varphi ), \quad  (r,\mathbold{\eta},\varphi) \in [0,1]\times [0,\pi]^{d-2}\times [0,2\pi],
\end{align} where $K\in \mathbb{N}$ again has to be determined. Now, let $r \in (0,1)$ be polynomial-time computable. Then, taking the estimates \eqref{estimate.addition} and \eqref{est.3} as well as Lemma \ref{le:polynome} into account, for all $(r,\mathbold{\eta},\varphi) \in [0,1]\times [0,\pi]^{d-2}\times [0,2\pi]$ and $N\in \mathbb{N}$, we obtain 
\begin{align*}
\vert u(r, \mathbold{\eta}, \varphi)-u_N(r, \mathbold{\eta}, \varphi)\vert &= \vert \sum_{\ell=N K+1}^{\infty} \sum_{m=1}^{N(d,l)} c_{\ell,m} r^\ell Y_{\ell,m} (\mathbold{\eta}, \varphi)\vert \\
&\leq \sum_{\ell=N K+1}^{\infty} \sum_{m=1}^{N(d,l)} \vert c_{\ell,m} \vert r^\ell \vert Y_{\ell,m} (\mathbold{\eta}, \varphi)\vert \\
&\leq  \sup_{l,m\in \mathbb{N}} \vert c_{\ell,m}\vert  \sum_{\ell=N K+1}^{\infty} r^\ell \sum_{m=1}^{N(d,l)} \vert Y_{\ell,m} (\mathbold{\eta}, \varphi )\vert\\
&\leq  \Vert g\Vert_{\infty} \left(\frac{2\pi^{\frac{d}{2}}}{\Gamma(\frac{d}{2})}\right)^{\frac{1}{2}} \sum_{\ell=N K+1}^{\infty} r^\ell N(d,l)^{\frac{1}{2}}\left(\frac{N(d,l)\Gamma(\frac{d}{2})}{2\pi^{\frac{d}{2}}}\right)^{\frac{1}{2}} \\
&\leq   \Vert g\Vert_{\infty}  \left(\frac{2\pi^{\frac{d}{2}}}{\Gamma(\frac{d}{2})}\right)^{\frac{1}{2}}  \sum_{\ell=N K+1}^{\infty} r^\ell N(d,l)^{\frac{1}{2}}\left(\frac{N(d,l)\Gamma(\frac{d}{2})}{2\pi^{\frac{d}{2}}}\right)^{\frac{1}{2}} \\
&=   \Vert g\Vert_{\infty} \sum_{\ell=N K+1}^{\infty} r^\ell  N(d,l)\\
&=  \Vert g\Vert_{\infty}  \sum_{\ell=N K+1}^{\infty} r^\ell \frac{(2l+d-2)}{l} {l+d-3\choose l-1}\\
&\leq \Vert g\Vert_{\infty}  \sum_{\ell=N K+1}^{\infty}  r^\ell  {l+d-2\choose l}\\
&\leq \frac{\Vert g\Vert_{\infty}}{(d-2)!}  \sum_{\ell=N K+1}^{\infty}  r^\ell  \frac{(l+d-2)!}{l!},\\
&= \frac{\Vert g\Vert_{\infty}}{(d-2)!}  \frac{r_0^{N K+1}}{(1-r_0)^{d-2}}P_{d-2}(r_0,N K+1)
\end{align*} for some polynomial $P_{d-2}$ of order $2(d-2)$. Again, since an exponentially decreasing function dominates any polynomial and $d$ is fixed, there exists a number $K\in \mathbb{N}$ such that 
\begin{align*}
    \vert u(r, \mathbold{\eta}, \varphi)-u_N(r, \mathbold{\eta}, \varphi)\vert < 2^{N} 
\end{align*} for all $(r,\mathbold{\eta}, \varphi)\in [0,r_0]\times [0,\pi]^{d-2}\times [0,2\pi]$ and $N\in \mathbb{N}$. \\\\
\textbf{Ad Upper Bound:} We wish to remark that the computation of $u_N$ requires the computation of $\sum_{\ell=0}^{N K} \sum_{m=1}^{N(d,l)} 1$, i.e., finitely many coefficients whose calculations are each in $\#P_1$. Therefore, the calculation of the map $(r,\mathbold{\eta},\varphi)\mapsto u(r,\mathbold{\eta},\varphi)$ is in $FP_1^{\#P_1}$ since the spherical harmonic functions are each polynomial-time computable. This reasoning shows that the solution operator maps maps polynomial-time computable input data to functions that have a computational complexity of at most $FP_1^{\#P_1}$. This shows an upper bound for the computational complexity. Next, we prove that there is also a lower bound for the complexity of the solution operator proving that the output function of the solution operator is in fact $\#P_1$-complete. Thus, if $FP_1={\#P_1}$, the $FP_1^{\#P_1}=FP_1$, and for every polynomial-time computable function $g$, the solution is polynomial-time computable.  \\\\
\textbf{Ad completeness:} Let $r_0 \in \mathbb{R}_c\cap [0 ,1)$, $\mathbold{\eta}_0 \in \mathbb{R}_c^{d-2}\cap [0 ,\pi]^{d-2}$ and $\varphi_0 \in \mathbb{R}_c\cap [0 ,2\pi)$ be polynomial-time computable numbers. Then, let $g_1\in\rmC([0,\pi]^{d-2})$ and $g_2\in\rmC([0,2\pi])$ are polynomial-time computable functions such that $g_2(0)=g_2(2\pi)$ and define
\begin{align*}
\tilde{g}(\mathbold{\eta},\varphi)=
       \frac{2}{\pi} g_1\left(\mathbold{\eta}\right)g_2\left(\varphi\right) \frac{ \vert T(r_0,\mathbold{\eta}_0,\varphi_0)- T(1,\mathbold{\eta},\varphi)\vert_2^d}{1-r_0^2}, \quad (\mathbold{\eta},\varphi) \in [0 ,\pi]^{d-2}\times [0,2\pi].
\end{align*} Now, we choose $g_1=1$ and $g_2$ on the interval $[0,1]$ according to Theorem \ref{Friedman1} and on the interval $[1,2\pi]$ let $g_2$ be a linear interpolation such that $g_2$ is continuous and periodic on $[0,2\pi]$. Hence, by choice the calculation of $\int_{0}^{2\pi} g_2(\varphi)\mathrm{d}\varphi$  is $\# P_1$-complete. 
This can be achieved by choosing, e.g., $g_1=1$ and $g_2$ on the interval $[0,1]$ according to Theorem \ref{Friedman1} and on the interval $[1,2\pi]$ a linear interpolation such that $g_2$ is continuous and periodic on $[0,2\pi]$. We note that $\tilde{g}$ is continuous and periodic in $\varphi$. From the construction of $\tilde{g}$, it follows that $\tilde{g}$ is also a polynomial-time computable function.
Then, for the solution, we obtain
	\begin{align*}
		u(r,\mathbold{\vartheta},\varphi)= \int_{0}^{2\pi} \int_{[0,\pi]^{d-2}} \frac{(1-r^2)\vert T(r_0,\mathbold{\vartheta}_0,\varphi_0)- T(1,\mathbold{\eta},\tau)\vert_2^d}{(1-r_0^2)\vert T(r,\mathbold{\vartheta},\varphi)- T(1,\mathbold{\eta},\tau)\vert_2^d}\frac{1}{2\pi^2} g_1(\mathbold{\eta})g_2(\tau) \mathrm{d}\tilde{S}(\mathbold{\eta}) \dd \tau 
	\end{align*}  for $r\in [0,1], \mathbold{\vartheta}\in [0,\pi]^{d-2}, \varphi\in [0,2\pi]$. Finally, for $(r,\mathbold{\vartheta},\varphi)=(r_0,\mathbold{\vartheta}_0,\varphi_0)$, we obtain
	\begin{align*}
		u(r_0,\mathbold{\vartheta}_0,\varphi_0) &=  \int_{0}^{2\pi} \int_{[0,\pi]^{d-2}} \frac{1}{2\pi^2} g_1(\mathbold{\eta})g_2(\tau) \dd \tilde{S}(\mathbold{\eta}) \dd \tau \\
  &=\frac{1}{2\pi^2} \int_{0}^{2\pi} g_2(\tau)\mathrm{d}\tau   \int_{[0,\pi]^{d-2}} 1 \dd \tilde{S}(\mathbold{\eta}) 
	\end{align*} Since $\frac{1}{2\pi^2}$ and $ \int_{[0,\pi]^{d-2}} 1 \dd \tilde{S}(\mathbold{\eta})$ are polynomial-time computable real numbers, the computation of the value $u(r_0,\mathbold{\vartheta}_0,\varphi_0)$ is $\# P_1$-complete. This shows also a lower bound for the complexity of the solution $u$. Hence, the computation of the solution operator $S$ is $\# P_1$-complete.
\end{proof}
Next, we present the proof of Theorem \ref{th:diffusion.compact}.
\begin{proof}[Proof of Theorem \ref{th:diffusion.compact}] \textbf{Ad Computability:} Since $x_L<\infty$, we can use Fourier series and write equivalently
\begin{align}
u(t,x)=\sum_{k=1}^\infty \mu_k \sin\left (\frac{\pi k x}{L}\right ) e^{-\frac{k^2\pi^2 \alpha t}{L^2}}, 
\end{align} where $\mu_k=\frac{2}{L}\int_0^L g(x) \sin\left (\frac{\pi k x}{L}\right ) \mathrm{d}x, k\in \mathbb{N}$.
Now, let $t_0>0$ be an arbitrarily but fixed polynomial-time computable real number and  $g$ be a continuous polynomial-time computable function.

Now, to $N\in \mathbb{N}$, we want to calculate the function $u_N$ given by 
\begin{align}
	u_N(t,x)=\sum_{k=1}^{N K} \mu_k \sin\left (\frac{\pi k x}{L}\right ) e^{-\frac{k^2\pi^2 \alpha t}{L^2}}, 
\end{align}  with $K\in \mathbb{N}$ to be determined 
such that
\begin{align}
    \vert u(t,x)-u_N(t,x)\vert <2^{-N} \quad \forall t\in [t_0,1], x\in [0,L].
\end{align} 
We note that by Theorem \ref{Friedman1}, the Fourier coefficients are polyomial-time computable if and only if $FP_1=\# P_1$ (see \textsc{Boche} und \textsc{Pohl} \cite{BocPoh21CBSA}). In order to calculate $u_N$, we need to calculate $K \cdot N$ times the Fourier coefficients, hence $u_N$ is computable in $FP_1^{\# P_1}$.  
We find
  \begin{align*}
  		\vert u(t,x)-u_N(t,x)\vert &= \vert \sum_{k=N K}^{\infty} \mu_k \sin\left (\frac{\pi k x}{L}\right ) e^{-\frac{k^2\pi^2 \alpha t}{L^2}} \vert \\
  		&\leq\sup_{k\in \mathbb{N}} \vert \mu_k\vert \sum_{k=N K}^{\infty}  e^{-\frac{k^2\pi^2 \alpha t}{L^2}}\\
    &\leq\sup_{k\in \mathbb{N}} \vert \mu_k\vert \sum_{k=N K}^{\infty}  e^{-\frac{k^2\pi^2 \alpha t_0}{L^2}}\\
  		&\leq2\Vert g\Vert_{\infty} \sum_{k=N K}^{\infty}  e^{-\frac{k\pi^2 \alpha t_0}{L^2}}\\
  		&\leq\frac{2\Vert g\Vert_{\infty}}{1-e^{-\frac{\pi^2 \alpha t}{L^2}}}(e^{-\frac{K\pi^2 \alpha t_0}{L^2}})^N
  \end{align*}
  
 Then, there exists $K\in \mathbb{N}$ such that 
 \begin{align}
		\vert u(t,x)-u_N(t,x)\vert < \frac{1}{2^N}, \quad \text{for all } (t,x)\in [t_0,\infty)\times [0,L], N\in \mathbb{N}.
	\end{align}
\textbf{Ad Upper Bound:} Hence, we found an algorithm that computes the function $u$ in $FP_1^{\# P_1}$ which is an upper complexity bound for the solution. Now, if $FP_1=\# P_1$, then $FP_1^{\# P_1}=FP_1$ and for every polynomial-time computable function $g$, the solution is also polynomial-time computable.\\
\newline 
\textbf{Ad Completeness:} Next, we want to show that the calculation is actually in $\# P_1$-complete. Therefore, we have to show that the solution has also a lower complexity bound that is in $\#P_1$ which in total shows $\# P_1$-completeness. To do so, let $x_0\in [0,L] $ be a polynomial-time computable real number.  
Therefore, choosing  $g^*(y)=\tilde{g}(y)e^{\frac{(y-x_0)^2}{4\pi \alpha t_0}} $, where $\tilde{g}(y)=\frac{1}{L}g(\frac{y}{L})$ in such a way that the computation of 
\begin{align}
\int_0^L \tilde{g}(y) \mathrm{d}y=\int_0^L\frac{1}{L}g\left(\frac{y}{L}\right) \mathrm{d}y=\int_0^1 g(y) \mathrm{d}y
\end{align} is $\# P_1$-complete.  

We obtain 
\begin{align}
	u(t,x) = \frac{1}{\sqrt{4\pi t}}\int_0^L e^{\frac{(y-x_0)^2}{4\pi t_0}-\frac{(y-x)^2}{4\pi t}}\tilde{g}(y) \mathrm{d}y
\end{align}  Now,  for $(t,x)=(t_0, x_0)$, there holds
\begin{align}
	u(t_0,x_0) = \frac{1}{\sqrt{4\pi t_0}}\int_0^L \tilde{g}(y) \mathrm{d}y
\end{align} Since $ \frac{1}{\sqrt{4\pi t^*}}$ y and $e^{-\frac{(\cdot-x_0)^2}{4\pi \alpha t_0}}$ are polynomial-time computable, the computation of $u(t_0,x_0)$ is $\# P_1$-complete. 
\end{proof}

Finally, the proof of Theorem \ref{th:diffusion.prl.f} is presented.
\begin{proof}[Proof of Theorem \ref{th:diffusion.prl.f}] \textbf{Ad Computability:} The proof of this theorem follows along the same lines as Theorem \ref{th:diffusion.prl.h}. First, we note that since the external force has a compact support in the space domain, it follows that for all $(t,x)\in [0,+\infty)\times [0,+\infty)$
\begin{align}\label{sol.heat.4}
u(t,x)&=\int_0^t\int_0^{y_0} \frac{1}{\sqrt{4\alpha \pi}}\left( \tilde{g}\left(t-s,\frac{y-x}{\sqrt{4\alpha}}\right)-\tilde{g}\left(t-s,\frac{y+x}{\sqrt{4\alpha}}\right)\right) f(y,s)\mathrm{d}y \mathrm{d}s \notag \\
&=\int_0^t\int_0^{y_0} \frac{1}{\sqrt{4\alpha \pi (t-s)}} \left( e^{-\frac{(y-x)^2}{4\alpha (t-s)}}-e^{-\frac{(y+x)^2}{4\alpha (t-s)}}\right) f(y,s)\mathrm{d}y \mathrm{d}s \\
&=\int_0^t\int_0^{y_0} \frac{1}{\sqrt{4\alpha \pi (t-s)}} e^{-\frac{(y-x)^2}{4\alpha (t-s)}} f(y,s)\mathrm{d}y \mathrm{d}s \notag \\
&\quad -\int_0^t\int_0^{y_0} \frac{1}{\sqrt{4\alpha \pi (t-s)}} e^{-\frac{(y+x)^2}{4\alpha (t-s)}} f(y,s)\mathrm{d}y \mathrm{d}s \notag \\
&= u^1(t,x)+u^2(t,x), \notag 
\end{align} where $\tilde{g}(t,x)=\frac{1}{t^{\frac{1}{2}}}e^{\frac{-x^2}{t}}=\frac{t}{x}g(t,x)$ with $g$ being defined in \eqref{function.g}. Next, with Corollary \ref{taylor.cor}, we obtain  
\begin{align}
\tilde{g}(t,y-x)-\tilde{g}(t,y+x)=\sum_{n=0}^\infty \frac{\tilde{g}^{(n)}(1,y-x)-\tilde{g}^{(n)}(1,y+x)}{n!}(t-1)^n 
\end{align} for $t\in (0,2), x,y\in  [0,+\infty)$ with $\tilde{g}^{(n)}$ being defined \eqref{derivative.2}. In addition, for each $x_0\in [0,+\infty)$, there exists some constant $C=C(x_0)>0$ such that  
\begin{align*} 
\frac{\tilde{g}^{(n)}(1,z)}{n!}\leq C(n+1) \quad \text{for all }z\in [0,x_0], n\in \mathbb{N}
\end{align*} which follows \eqref{derivative.bound.g2}. Then, similar to before, we define for $N\in \mathbb{N}$ the sequence of polynomial-time computable functions
\begin{align*}
    u_N(t,x)&=\sum_{n=0}^{N^3} \int_0^{t-\frac{1}{N}} \int_0^{y_0}\frac{\tilde{g}^{(n)}(1,\frac{y-x}{\sqrt{4\alpha}})-\tilde{g}^{(n)}(1,\frac{y+x}{\sqrt{4\alpha}}) }{\sqrt{4\alpha \pi} n!}   (t-s-1)^n f(y,s)\mathrm{d}y \mathrm{d}s, \\
    &=\sum_{n=0}^{N^3} \int_0^{t-\frac{1}{N}} \int_0^{y_0}\frac{\tilde{g}^{(n)}(1,\frac{y-x}{\sqrt{4\alpha}}) }{\sqrt{4\alpha \pi} n!}  (t-s-1)^n f(y,s)\mathrm{d}y \mathrm{d}s\\
    &\quad- \sum_{n=0}^{N^3} \int_0^{t-\frac{1}{N}} \int_0^{y_0}\frac{\tilde{g}^{(n)}(1,\frac{y+x}{\sqrt{4\alpha}}) }{\sqrt{4\alpha \pi} n!}  (t-s-1)^n f(y,s)\mathrm{d}y \mathrm{d}s\\
    &= u_N^1(t,x)+u_N^2(t,x)
\end{align*} for all $t\in [\frac{1}{N},1]$ and $u_N(t,x)$=0 for all $t\in [0,\frac{1}{N})$. Now, we want to show that there exists some $N_0$ such that  
\begin{align*} 
    \vert u(t,x)-u_N(t,x)\vert &=  \vert u^1(t,x)-u^1_N(t,x)+u^2(t,x)-u^2_N(t,x)\vert \\
    &\leq \vert u^1(t,x)-u^1_N(t,x)\vert +\vert u^2(t,x)-u^2_N(t,x)\vert \\
    &\leq 2^{-(N+1)}+2^{-(N+1)}=2^{-N} 
\end{align*} for all $t\in [0,1]$ and $N\geq N_0$. For simplicity, we only show the estimate involving the terms $u^1$ and $u_N^1$ since the estimation of the other term is analogous. To do so, let $t\in [\frac{1}{N},1]$. Then, 
\begin{align*}
   & \vert u^1(t,x)-u^1_N(t,x)\vert \\
   &=  \big \vert \int_0^t\int_0^{y_0} \frac{1}{\sqrt{4\alpha \pi (t-s)}} e^{-\frac{(y-x)^2}{4\alpha (t-s)}} f(y,s)\mathrm{d}y \mathrm{d}s\\ 
   &\quad -  \sum_{n=0}^{N^3} \int_0^{t-\frac{1}{N}} \int_0^{y_0}\frac{\tilde{g}^{(n)}(1,\frac{y-x}{\sqrt{4\alpha}}) }{\sqrt{4\alpha \pi} n!}  (t-s-1)^n f(y,s) \mathrm{d}y\mathrm{d}s \big \vert\\
   &=  \big \vert \int_{t-\frac{1}{N}}^{t}\int_0^{y_0} \frac{1}{\sqrt{4\alpha \pi (t-s)}} e^{-\frac{(y-x)^2}{4\alpha (t-s)}} f(y,s)\mathrm{d}y \mathrm{d}s\\
  & \quad + \int_0^{t-\frac{1}{N}}\int_0^{y_0} \frac{1}{\sqrt{4\alpha \pi (t-s)}} e^{-\frac{(y-x)^2}{4\alpha (t-s)}} f(y,s)\mathrm{d}y \mathrm{d}s\\
   &\quad -  \sum_{n=0}^{N^3} \int_0^{t-\frac{1}{N}} \int_0^{y_0}\frac{\tilde{g}^{(n)}(1,\frac{y-x}{\sqrt{4\alpha}}) }{\sqrt{4\alpha \pi} n!} (t-s-1)^n f(y,s)\mathrm{d}y \mathrm{d}s \big \vert \\
  &\leq    \big \vert \int_{t-\frac{1}{N}}^{t}\int_0^{y_0} \frac{1}{\sqrt{4\alpha \pi (t-s)}} e^{-\frac{(y-x)^2}{4\alpha (t-s)}} f(y,s)\mathrm{d}y \mathrm{d}s \big \vert  \\
   &\quad +  \big \vert \sum_{n=N^3+1}^{\infty} \int_0^{t-\frac{1}{N}} \int_0^{y_0}\frac{\tilde{g}^{(n)}(1,\frac{y-x}{\sqrt{4\alpha}}) }{\sqrt{4\alpha \pi} n!} (t-s-1)^n f(y,s)\mathrm{d}y \mathrm{d}s \big \vert\\
   &= I_1+I_2.
\end{align*}  
Again, we estimate terms $I_1$ and $I_2$ separately. As in the previous case, we make use of the fact that the $t\mapsto \frac{1}{\sqrt{4\alpha \pi t}} e^{-\frac{x^2}{4\alpha t}}$ is monotonically increasing on $[0,\frac{(x-y)^2}{6\alpha}]$. Since, $x\in[x_0,x_1], $ and $y\in [0,y_0]$ with $y_0<x_0$ , we can choose $\tilde{N}_0=\tilde{N}_0(x,\alpha)\in \mathbb{N}$ such that $\frac{1}{\tilde{N}_0}\leq \frac{(x_0-y_0)^2}{6\alpha}$. Then, for all $N\geq \tilde{N}_0$, we obtain 
\begin{align*}
   I_1 &\leq    \int_{t-\frac{1}{N}}^{t}\int_0^{y_0} \frac{1}{\sqrt{4\alpha \pi (t-s)}} e^{-\frac{(y-x)^2}{4\alpha (t-s)}} \vert f(y,s)\vert \mathrm{d}y \mathrm{d}s  \\
   &=\int_{0}^{\frac{1}{N}}\int_0^{y_0} \frac{1}{\sqrt{4\alpha \pi s}} e^{-\frac{(y-x)^2}{4\alpha s}} \vert f(y,t-s)\vert \mathrm{d}y \mathrm{d}s  \\
   &\leq \int_{0}^{\frac{1}{N}}\frac{1}{\sqrt{4\alpha \pi s}} e^{-\frac{(y_0-x_0)^2}{4\alpha s}} \int_0^{y_0} \vert f(y,t-s)\vert \mathrm{d}y \mathrm{d}s  \\
   &\leq \int_{0}^{\frac{1}{N}} \frac{N^{\frac{1}{2}}}{\sqrt{4\alpha \pi }} e^{-\frac{(y_0-x_0)^2 N}{4\alpha}} \int_0^{y_0}\vert f(y,t-s)\vert \mathrm{d}y \mathrm{d}s  \\
    &\leq  \frac{y_0 N^{-\frac{1}{2}}}{\sqrt{4\alpha \pi }} e^{-\frac{(y_0-x_0)^2 N}{4\alpha}} \Vert f\Vert_{\rmC([0,2]\times [0,y_0])} \\
    &\leq  \frac{y_0}{\sqrt{4\alpha \pi }} e^{-\frac{(y_0-x_0)^2 N}{4\alpha}} \Vert f\Vert_{\rmC([0,2]\times [0,y_0])}.
\end{align*} Now, we replace $N$ by $K\times N$ and choose $K\in \mathbb{N}$ such that there exists some $N_1=N_1(x,\alpha,h)\in \mathbb{N}$ such that $I_1\leq 2^{-N}$ for all $N\geq N_1$. For the term $I_2$, we make use of the estimate  \eqref{derivative.bound.g2}. Hence, there exists some constant $C>0$ such that 
\begin{align*}
   I_2 &\leq   \sum_{n=N^3+1}^{\infty} \int_0^{t-\frac{1}{N}} \int_0^{y_0}\frac{\tilde{g}^{(n)}(1,\frac{y-x}{\sqrt{4\alpha}}) }{\sqrt{4\alpha \pi} n!} \vert t-s-1\vert ^n \vert f(y,s)\vert \mathrm{d}y \mathrm{d}s \\
   &\leq   \frac{C}{\sqrt{4\alpha \pi}} \sum_{n=N^3+1}^{\infty} \int_0^{t-\frac{1}{N}} (n+1) \vert t-s-1\vert ^n \int_0^{y_0} \vert f(y,s)\vert \mathrm{d}y \mathrm{d}s \\
   &=   \frac{C}{\sqrt{4\alpha \pi}} \sum_{n=N^3+1}^{\infty} (n+1) \int_{\frac{1}{N}}^t \vert s-1\vert ^n \int_0^{y_0} \vert f(y,t-s)\vert \mathrm{d}y \mathrm{d}s \\
    &\leq   \frac{C}{\sqrt{4\alpha \pi}} \sum_{n=N^3+1}^{\infty} (n+1)  \left (1-\frac{1}{N}\right)^n  \int_{\frac{1}{N}}^t\int_0^{y_0} \vert f(y,t-s)\vert \mathrm{d}y \mathrm{d}s \\
     &\leq   \frac{C 2y_0}{\sqrt{4\alpha \pi}} \sum_{n=N^3+1}^{\infty} (n+1)  \left (1-\frac{1}{N}\right)^n  \Vert f\Vert_{\rmC([0,2]\times [0,y_0])} \\ 
    &=  \frac{C y_0}{\sqrt{\alpha \pi}} \Vert f\Vert_{\rmC([0,2]\times [0,y_0])}  \sum_{n=N^3+1}^{\infty} (n+1) \left (1-\frac{1}{N}\right )^n \\
    &= \frac{C y_0}{\sqrt{\alpha \pi}} \Vert f\Vert_{\rmC([0,2]\times [0,y_0])}  \left (1-\frac{1}{N}\right )^{N^3+1} \left( N^4-N^5+N^2\right)  \\
    &\leq \frac{C y_0}{\sqrt{\alpha \pi}} \Vert f\Vert_{\rmC([0,2]\times [0,y_0])}\left (1-\frac{1}{N}\right )^{N^3+1}  (N^2+1)^2  \\
    &\leq\frac{C y_0}{\sqrt{\alpha \pi}} \Vert f\Vert_{\rmC([0,2]\times [0,y_0])}\left (1-\frac{1}{N}\right )^{N\cdot N^2} (N^2+1)^2\\
    &\leq  \frac{C y_0}{\sqrt{\alpha \pi}} \Vert f\Vert_{\rmC([0,2]\times [0,y_0])}e^{- N^2} (N^2+1)^2,
\end{align*}
where we used again the formula for the arithmetico-geometric series \eqref{arit.geom} and the fact that $\left (1-\frac{1}{N}\right )^N$ is strictly increasing in $N$ and converges to $e^{-1}$ from below as $N\rightarrow \infty$. 
Hence, there exists another $\tilde{N}_2=\tilde{N}_2(x_0,y_0,\alpha,h)\in \mathbb{N}$ independently of $t$ such that $I_2\leq 2^{-N}$ for all $N\geq \tilde{N}_2$.
Finally, for $t\in [0,\frac{1}{N}]$, there holds 
\begin{align*}
   \vert u^1(t,x)-u^1_N(t,x)\vert &=  \vert u^1(t,x)\vert \\
   &=\big \vert \int_0^t\int_0^{y_0} \frac{1}{\sqrt{4\alpha \pi (t-s)}} e^{-\frac{(y-x)^2}{4\alpha (t-s)}} f(y,s)\mathrm{d}y \mathrm{d}s\big \vert \\
   &\leq  \int_0^t\int_0^{y_0} \frac{1}{\sqrt{4\alpha \pi (t-s)}} e^{-\frac{(y-x)^2}{4\alpha (t-s)}} \vert f(y,s)\vert 
 \mathrm{d}y \mathrm{d}s \\
   &= \int_0^t \int_0^{y_0} \frac{1}{\sqrt{4\alpha \pi s}} e^{-\frac{(y-x)^2}{4\alpha s}} \vert f(y,t-s)\vert \mathrm{d}y \mathrm{d}s\\
   &\leq  \int_0^{\frac{1}{N}}\int_0^{y_0} \frac{1}{\sqrt{4\alpha \pi s}} e^{-\frac{(y-x)^2}{4\alpha s}} \vert f(y,t-s)\vert \mathrm{d}y \mathrm{d}s\\
    &\leq  \int_0^{\frac{1}{N}}\int_0^{y_0} \frac{1}{\sqrt{4\alpha \pi s}} e^{-\frac{(y_0-x_0)^2}{4\alpha s}} \vert f(y,t-s)\vert \mathrm{d}y \mathrm{d}s\\
    &\leq \frac{N^{\frac{1}{2}}}{\sqrt{4\alpha \pi }} e^{-\frac{(y_0-x_0)^2 N}{4\alpha}} \int_{0}^{\frac{1}{N}} \int_0^{y_0}\vert f(y,t-s)\vert \mathrm{d}y \mathrm{d}s  \\
    &\leq  \frac{y_0 N^{-\frac{1}{2}}}{\sqrt{4\alpha \pi }} e^{-\frac{(y_0-x_0)^2 N}{4\alpha}} \Vert f\Vert_{\rmC([0,2]\times [0,y_0])} \\
    &\leq  \frac{y_0}{\sqrt{4\alpha \pi }} e^{-\frac{(y_0-x_0)^2 N}{4\alpha}} \Vert f\Vert_{\rmC([0,2]\times [0,y_0])} 
\end{align*} where we again made use of the fact that the mapping $t\mapsto \frac{x}{\sqrt{4\alpha \pi t^3}} e^{-\frac{x^2}{4\alpha t}}$ is strictly increasing on $[0,\frac{x}{\sqrt{6\alpha}}]$ for $N$ sufficiently large, i.e., $\frac{1}{N}\leq \frac{x^2}{6\alpha}$. As before, there exists another $\tilde{N}_3=\tilde{N}_2(x_0,y_0,\alpha,h)\in \mathbb{N}$ independently of $t$ such that $\vert u^1(t,x)\vert\leq 2^{-N+1}$ for all $N\geq \tilde{N}_3$. Choosing $N_0=\max_{i=1,2,3}\tilde{N}_i$, there holds
\begin{align*} 
    \vert u(t,x)-u_N(t,x)\vert \leq  2^{-N} \quad \text{for all }(t,x)\in [0,1]\times  [x_0,x_1] \text{ and }N\geq N_0,
\end{align*} which completes the proof. As mentioned before, the calculation of the term involving $u^2$ is done analogously. \\\\
\textbf{Ad Upper Bound:} Since the calculation of $u_N$ requires a finite number of calculations in which each being computable in $\#P$, the computation of $u_N$ has a computational complexity of $FP^{\#P}$. Thus, if $FP=\#P$, then $FP^{\#P}=FP$, and for every polynomial-time computable function $f$, the solution is polynomial-time computable.
\end{proof}

\section*{Acknowledgments}
This work of G. Kutyniok and H. Boche was supported in part by the ONE Munich Strategy Forum (LMU Munich, TU Munich, and the Bavarian Ministery for Science and Art).

G. Kutyniok acknowledges support from the Konrad Zuse School of Excellence in Reliable AI (DAAD), the Munich Center for Machine Learning (BMBF) as well as the German Research Foundation under Grants DFG-SPP-2298, KU 1446/31-1 and KU 1446/32-1 and under Grant DFG-SFB/TR 109, Project C09 and the Federal Ministry of Education and Research under Grant MaGriDo.

This work of H. Boche was supported in part by the German Federal Ministry of Education
and Research (BMBF) under Grant 16ME0442.

\bibliographystyle{my_alpha}
\bibliography{bib_aras}

%
%

%

\end{document}